\documentclass[12pt]{iopart}
\usepackage{iopams}

\usepackage{amssymb}
\usepackage{graphicx}
\usepackage{bm}
\usepackage{color}
\usepackage{subfig}

\newcommand*{\defeq}{\mathrel{\vcenter{\baselineskip0.5ex \lineskiplimit0pt
                     \hbox{\scriptsize.}\hbox{\scriptsize.}}}
                     =}
\newcommand{\beq}{\begin{equation}}
\newcommand{\eeq}{\end{equation}}
\newcommand{\beqar}{\begin{eqnarray}}
\newcommand{\eeqar}{\end{eqnarray}}
\newcommand{\bea}{\begin{eqnarray}}
\newcommand{\eea}{\end{eqnarray}}
\newcommand{\bcen}{\begin{center}}
\newcommand{\ecen}{\end{center}}

\newcommand{\bra}[1]{\left< #1 \right|}
\newcommand{\ket}[1]{\left| #1 \right>}
\newcommand{\ketbra}[2]{\left| #1 \right> \left< #2 \right|}
\newcommand{\braket}[2]{\left< #1 \vert #2 \right>}

\newcommand{\ave}[1]{\left< #1 \right>}

\newcommand{\half}{\frac{1}{2}}
\newcommand{\Hop}{\hat H}

\newcommand{\Xopb}{{\hat X}}
\newcommand{\rop}{\hat\rho}

\newcommand{\sop}{\hat{\sigma}}

\newcommand{\pop}{\hat{p}}
\newcommand{\qop}{\hat{q}}
\newcommand{\Iop}{\hat{I}}
\newcommand{\Top}{\hat{T}}

\begin{document}

\title{Noise Resistant Quantum Control Using Dynamical Invariants}

\author{Amikam Levy$^1$, A. Kiely$^2$, J. G. Muga$^2$, R. Kosloff$^1$, and E. Torrontegui$^{1,3}$}

\address{$^1$ Institute of Chemistry and The Fritz Haber Research Center, The Hebrew University, Jerusalem 91904, Israel}
\address{$^2$ Departamento de Qu\'{\i}mica F\'{\i}sica, Universidad del Pa\'{\i}s Vasco - Euskal Herriko Unibertsitatea, Apdo. 644, Bilbao, Spain}
\address{$^3$ Instituto de F\'{\i}sica Fundamental IFF-CSIC, Calle Serrano 113b, 28006 Madrid, Spain}

\eads{\mailto{amikamlevy@gmail.com}, \mailto{eriktm@iff.csic.es}}


\begin{abstract}
A systematic approach to design robust control protocols against the influence of different types of noise is introduced.
We present control schemes which protect the decay of the populations avoiding dissipation in the adiabatic and non-adiabatic regimes and minimize the effect of dephasing.
The effectiveness of the protocols is demonstrated in two different systems. 
Firstly we present the case of population inversion of a two level system in the presence of either one or two simultaneous noise sources. 
Secondly, we present an example of the expansion of coherent and thermal states in harmonic traps, subject to noise arising from monitoring and modulation of the control respectively.

\end{abstract}

\section{Introduction}
A major obstacle to manipulate quantum systems and develop quantum technologies is the unavoidable presence of noise.
There are different types of noise sources that disturb control protocols, including noise induced by the environment, noise caused from interaction with the measurement apparatus, or errors in the implementation of the control protocol. 
Different approaches proposed to suppress or mitigate the effects of noise include (for a recent review  see \cite{koch16}): the use of decoherence-free subspaces which
are immune to noise \cite{lidar03}, correction of errors using quantum feedback controls \cite{wisemanbook}, performing sudden interactions on the system on timescales for which the noise only slightly interferes with the process such as in dynamical decoupling \cite{lloyd99}, and  applying shortcuts to adiabaticity (STA)\cite{Chen2010, torrontegui13}.
The noise has also been proposed as a resource to achieve the desired control in specific processes \cite{levy12,levy17,koch15b}.

A large family of control protocols are based on adiabatic following of instantaneous eigenstates of a time dependent Hamiltonian by smoothly
and slowly changing control parameters. These methods are widespread as they are in principle robust 
against control imperfections. However, they are also prone to suffer the effects of noise due to long operation times.   
As a result the fidelity of the final state with respect to the target state is reduced \cite{koch16, khasin11}.  

Shortcuts to adiabaticity are control methods to derive protocols which reach fidelities of slow adiabatic processes in significantly shorter times. 
STA have been applied  in a wide variety of fields including 
quantum computation \cite{sarandy11, Palmero2017}, cooling \cite{Chen2010, Onofrio2017}, quantum transport \cite{Torrontegui2011,bowler}, 
quantum state preparation \cite{muga12b, 2levelEXP1, 2levelEXP2, Zhou2017_330}, cold atoms manipulation \cite{Torrontegui2012,sta_exp1,schaff11,split_sta, Kiely2016}, 
many-body state engineering \cite{adolfo_multi, SebCriAdo, Takahashi2013, Martinez-Garaot2015_043406} and polyatomic molecules control \cite{Masuda2015}, design of optical devices \cite{Tseng2012, Martinez-Garaot2014_2306} and linear chains \cite{Longhi2017}, or 
mechanical engineering \cite{Torrontegui2017, Gonzalez2017}. 
\par
STA provide  a strategy to combat the effects of noise thanks to two different mechanisms: (i) In principle faster than adiabatic processes are desirable to avoid pervasive, long interactions with the environment.
In practice the fidelities  may present maxima at specific times \cite{levy17}, and  STA can be set for these optimally short process times. 
Many studies have tested the achieved robustness with master equations including the noise, see e.g.  \cite{Chen2014, Chen2015_012325}; (ii) in addition, the parameter paths leading to STA are typically not unique, so this freedom may be used to choose the most robust ones with respect to specific perturbations, noise or control imperfections. 
This optimization has been performed so far by minimizing the excitation energy or maximizing the fidelity in perturbative schemes, e.g. in two-level systems \cite{Ruschhaupt2012,Lu2013,Daems2013, Kiely2014} or for ion transport \cite{Lu2014_063414}, using decoherence free subspaces \cite{Wu2015, Wu2017_042104}, super-operator
\cite{Sarandy2007} and
non-Hermitian invariants \cite{Maamache2017}, and effective Hamiltonians \cite{Luo2015} . 

%

In this work we introduce an alternative  systematic method for smooth control under the influence of noise which is applicable for both  adiabatic and nonadiabatic  time scales. 
%
%
%
The technique proposed intends to go beyond the perturbative regime and can be applied to the strong noise regime \cite{Kiely2017}.  
The central idea is to inverse engineer the noiseless Hamiltonian by designing its  
dynamical invariants (i.e., the dynamics of the noiseless system) such that the noise has a minimal effect.
The control functions to be minimized are state independent, and measure the deviation of the actual invariant, i.e., the one for the full dynamics including noise, from the noiseless invariant.     
This technique does not require one to solve the full dynamics iteratively as is often done in optimal control methods \cite{rabitz88,koch15}. 
This property makes the method appealing and simple for implementation.
Furthermore, it is not restricted to very fast operations, where tipycally very short control time is limited by experimental constraints.
\par 
The dynamics of the system including the dissipative term resulting from the noise  takes the form (for $\hbar=1$)
\beq
\label{eq:master_equation}
\frac{d \rop}{dt}=-i[\Hop(t),\rop]+\mathcal{L}\rop,
\eeq
where
\beq
\label{eq:dissipator}
\mathcal{L}\rop=-\sum_{ k} \eta_{ k} [\Xopb_{k} (t),[\Xopb_k (t),\rop]]\quad $with$ \quad \eta_k > 0.
\eeq 
Here $\Hop(t)$ is the total Hamiltonian of the system including the control Hamiltonian and 
the noise term given by Eq. (\ref{eq:dissipator}). 
The ${\Xopb}_k(t)$ represent Hermitian operators acting on the Hilbert space of the system and can be explicitly  time dependent.
The pre-factors  $\eta_k$ are scaling factors representing the strength of the noise and may have different dimension depending on ${\Xopb}_k(t)$. The sum over $k$ 
includes the possibility of independent types of noise simultaneously affecting the dynamics.
This equation was derived in different  contexts, including the singular coupling limit \cite{gorini76}, phase noise \cite{milburn91}, action noise \cite{levy17}, amplitude noise \cite{kallush14},  noise from monitoring weakly some quadrature of the system \cite{wisemanbook, diosi88}, Gaussian noise and Poisson noise for SU(2) algebra \cite{Luczka1991, levy14a}, and more \cite{breuer,luczka91}.
\par 
In Sec. \ref{sec:DI} we present the main results of this work. 
First we construct the dynamical invariant method in the density operator formalism which can also be then applied to the study of noise and naturally extend the treatment from pure states to general mixed states. 
We derive two measures to quantify the effect of noise introducing constraints on the noiseless dynamical invariant. 
In Sec. \ref{sec:two_level} we study the example of the two-level system with single and multiple noise terms in the dynamics. 
Section \ref{sec:ho} studies  the control of thermal and coherent states of the harmonic oscillator.
We conclude with a discussion and outlook on future work in Sec. \ref{sec:discussion}, plus some technical appendices. 
\section{Dynamical Invariant and noise resistant control}
\label{sec:DI}
\subsection{Dynamical Invariant for Unitary Dynamics}
%
%
%
We refer the reader who is unfamiliar with the dynamical invariant method  to \ref{app:general} where we present the method in the wave function formalism. 
\par
For noiseless, unitary dynamics, the evolution of the density operator is described by
\beq
\frac{d \rop}{dt}=-i[\Hop(t),\rop].
\eeq 
A  dynamical invariant satisfies the equation \cite{Lewis1969}
\beq
\label{eq:inv_dyn}
i\frac{\partial \hat{I}(t)}{\partial t}-[\Hop(t), \hat{I}(t)]=0,
\eeq
and can be expressed in diagonal form,   
\beq
\label{eq:inv}
\hat{I}(t)=\sum_k \lambda_k\ketbra{\phi_k}{\phi_k}.
\eeq
Here $\lambda_k$ are the real time independent eigenvalues and $\ket{\phi_k} \equiv \ket{\phi_k(t)}$ are the time dependent eigenvectors of the invariant.
In this basis, the density matrix elements $\rho_{lk} \equiv \bra{\phi_l}\rop(t)\ket{\phi_k} $ can be calculated from
\beq
\label{eq:rho_u}
\dot{\rho}_{lk}=i\left(\bra{\phi_l}i\frac{\partial}{\partial t} -\Hop(t) \ket{\phi_l}-\bra{\phi_k}i\frac{\partial}{\partial t} -\Hop(t) \ket{\phi_k} \right)\rho_{lk},
\eeq
where the dot represents the time derivative.
The off diagonal terms of the density matrix depend on the difference of time derivatives of two Lewis-Riesenfeld phases (compare with Eq.(\ref{appeq:phase})), while the populations remain constant with time \cite{Lewis1969}.
As the system is driven through the instantaneous eigenstates of the invariant,
imposing $[\hat{I}(0),\Hop(0)]=[\hat{I}(t_f),\Hop(t_f)]=0$ we ensure that the system starts and ends in an energy eigenstate of $\hat H$ without unwanted excitations. 
The state transfer is designed by choosing $\hat I(t)$ and then determining $\hat H(t)$. (See \ref{app:general} for more details.)
\subsection{Dynamical invariant under the influence of noise} 
We now consider the influence of Eq. (\ref{eq:dissipator}) on the control process.
In order to demonstrate the effect of noise we consider a single operator $\Xopb \equiv \Xopb_1(t)$ and $\eta\equiv\eta_1$.
In a later example, we will also consider the case for simultaneous noise sources. 
The dynamics for an arbitrary observable $\hat{A}$ including the noise effect in the Heisenberg representation reads
\beq
\label{eq:me_heisen}
\frac{d\hat{A}}{dt} = \frac{\partial \hat{A}}{\partial t}+i[\hat H,\hat{A}]-\eta[ \Xopb,[\Xopb,\hat{A}]].
\eeq
Assuming the structure of the invariant  (\ref{eq:inv}) for the unitary dynamics we insert it in Eq. (\ref{eq:me_heisen}) to account for the noise.
The eigenvalues of the invariant $\lambda_l$ are now  no longer constant in time and evolve according to  
\beq
\label{eq:lambdadot}
\dot{\lambda}_l=2\eta\left(\lambda_l\langle{\phi_l}|\Xopb^2|{\phi_l}\rangle-\sum_k \lambda_k|\langle{\phi_k}|\Xopb|{\phi_l}\rangle|^2\right).
\eeq 
Note that if $\{ |\phi_k\rangle\}$ are eigenstates of $\Xopb$ then  $\dot{\lambda}_l=0$ as required in the unitary noiseless method.
In this case the invariant is not affected by the noise. 
Although the requirement for $\Iop$ and $\Xopb$ having common eigenstates cannot generally be achieved for all times during the process, the effect of
noise can be significantly  reduced by constructing the Hamiltonian from an invariant which shares common eigenvectors with those of $\Xopb$ during most of the process.
Since at final time we impose that the invariant and the Hamiltonian share common eigenstates, protecting the invariant from the noise will drive the system to the desired  target state.
\par
To express the density matrix elements, Eq. (\ref{eq:rho_u}) is now modified by adding to the r.h.s. the additional term 
$\dot{\rho}_{lk}^d\equiv \langle\phi_l|\mathcal{L}\rop|\phi_k\rangle$  which accounts for dissipation and decoherence resulting from the noise term and is given by
\beqar
\dot{\rho}_{lk}^d &=&-\eta\bigg[\sum_n( \rho_{nk}\langle\phi_l|\Xopb^2 |\phi_n\rangle +\rho_{ln}\langle\phi_n|\Xopb^2| \phi_k\rangle) \\ \nonumber
&-&2\sum_{nm}\rho_{nm}\langle\phi_l|\Xopb |\phi_n\rangle\langle\phi_m|\Xopb |\phi_k\rangle\bigg].
\eeqar
In the limit that $\Iop$ and $\Xopb$ share common eigenstates the contribution due to noise to the change in population and the off diagonal terms is
\beqar
\label{eq:rho_limits}
\dot{\rho}_{kk}^d & \rightarrow & 0
\\ \nonumber
\dot{\rho}_{lk}^d & \rightarrow & -\eta \left(x_l-x_k\right)^2\rho_{lk},
\eeqar  
where $\left\lbrace x_k \right\rbrace$ are the eigenvalues of $\Xopb$. Note that in this limit the decay of the populations in the invariant eigenbasis is suppressed, however, a decay of
the coherences is still present, although it can be minimized for sufficiently fast 
processes.  

The strategy proposed here relies on the ample freedom provided by STA. 
By adding constraints on the unitary invariant we can design a control Hamiltonian that optimizes the fidelity under the influence of the noise.

To identify the amount of overlap between the two bases sets of $\hat I$ and $\Xopb$ we define the overlap matrix $S$ with the entries
\beq
S_{ij}(t)=\braket{\phi_i}{\psi_j}.
\eeq 
Here $\left\lbrace \ket{\psi_j}\right\rbrace$ are the eigenvectors of $\Xopb$.
The sum of the overlap matrix can be bounded by: $n \leqslant \sum_{ij}^n\vert S_{ij} \vert < n^2$, 
where $n$ is the dimension of the space. The upper bound is not tight, and obtaining a tight bound, typically, becomes a difficult optimization problem for high dimension. However, 
we are only interested in minimizing $S$. In the scenario of $n=2$, the tight upper bound is given by $2\sqrt{2}$.
Next, we define the measure for the overlap along the process as the time average of the distance between the overlap matrix and its minimal value.
\beq
\label{eq:overlap}
\mathcal{O} \defeq \frac{1}{t_f}\int_0^{t_f}\left(\sum_{ij}^n\vert S_{ij} \vert - n \right) dt. 
\eeq
The measure $\mathcal{O}$ is zero if and only if the eigenbasis of $\Xopb$ and $\Iop$ are identical.  
A different measure which stems from similar considerations and in some cases can be easier to compute is 
\beq
\label{eq:commu_norm}
\mathcal{A} \defeq z^{-1}\int_0^{t_f}\Vert[\Xopb(t),\hat{I}(t)] \Vert dt
=\frac{\int_0^{t_f} dt \sqrt{2 \mbox{tr}(\Iop^2 \Xopb^2)-2 \mbox{tr} (\Iop \Xopb \Iop \Xopb)}}{2\int_0^{t_f}dt \sqrt{\mbox{tr}(\Iop^2 \Xopb^2)} }.
\eeq
In  the above expressions we use the Frobenius norm defined as $\Vert \mathcal{M} \Vert \equiv \sqrt{\tr\left(\mathcal{M}\mathcal{M}^{\dagger} \right)}$. 
The normalization factor
 $z=2\int_0^{t_f}\Vert\Xopb(t)  \Iop(t) \Vert dt$  guarantees that $\mathcal{A}$ is dimensionless and equal or smaller than 1 (this is an immediate consequence of the sub-additivity of the Frobenius norm).   
The measure $\mathcal{A}$ is zero if and only if $\Xopb$ and $\hat{I}$ commute at all times during the process
$\Iop\Xopb=\Xopb\Iop$, and takes the maximal value $\mathcal{A}= 1$ when $\Iop\Xopb=-\Xopb\Iop$.
For unbounded operators  extra care is needed. 
The norm should be calculated on a finite domain or using other techniques as will be demonstrated in a later section. 

In order to improve the fidelity of the evolved state with respect to the target by
minimizing the effect of noise, the controls that drive the system are inverse engineered through the
 invariant $\hat I(t)$ of the unitary dynamics subject to the minimization of the measures $\mathcal{O}$ or $\mathcal{A}$.

\section{Two-level system}
\label{sec:two_level}
As a first example we consider the control problem of a full population inversion in a two level system (TLS) \cite{muga12b,muga12,demirplak03,berry09,rabitz08}. The Hamiltonian takes the form:
\beq
\label{H2}
\Hop(t)=\frac{ \Delta(t)}{2}\sop_z + \frac{\Omega(t)}{2}  \sop_x,
\eeq
where $\Delta(t)$ and  $\Omega(t)$ are real, time-dependent  functions resulting from an interaction with some external field, and $\sop_z$ and $\sop_x$ are the Pauli matrices. 
Initially the system is set to the ground state, $\rop_{0}=\ketbra{0}{0}$, with the initial Hamiltonian  corresponding to $\Delta(0)=\Delta_0$ and $\Omega(0)=0$. The desired target state  $\rop_{tar}=\ketbra{1}{1}$ corresponds to the ground state of the final Hamiltonian $\Delta(t_f)=-\Delta_0$ and $\Omega(t_f)=0$.
To evaluate the success of the control protocol we will use the fidelity
\beq
\label{fid_def}
\mathcal{F}\defeq \tr\sqrt{\sqrt{\rop(t_f)}\rop_{tar}\sqrt{\rop(t_f)}},
\eeq
that measures the overlap between the final state and the target state $\rop_{tar}$. 
To connect the states $\hat\rho(0)$ and $\hat\rho_{tar}$ we engineer the controls $\Delta(t)$ and $\Omega(t)$ from the dynamical invariant.
Associated with the Hamiltonian (\ref{H2}) there is a dynamical invariant expressed as (see \ref{app:su2}),
\beq
\label{Isu2}
\Iop(t)= \sqrt{\Delta^2(0) + \Omega^2(0)}
\left(\begin{array}{cc} \cos(G) & \sin(G)e^{iB} \\ \sin(G)e^{-iB} & -\cos(G) \end{array}\right),
\eeq
where $G\equiv G(t)$ and $B\equiv B(t)$ are auxiliary real time dependent functions of the invariant obeying 
\beq
\label{DelOm}
\Delta=-\dot B+\frac{\dot G}{\tan (G)\tan (B)}, \quad \Omega=\frac{\dot G}{\sin (B)}.
\eeq
The frictionless conditions $[\hat H(t_b),\hat I(t_b)]=0$ at $t_b=0, t_f$ impose at the boundary times fix
$G(0)=\pi$, $G(t_f)=0$, $\dot G(t_b)=0$, leaving $B(t_b)$ and $\dot B(t_b)$ as free parameters (\ref{app:su2}).
At intermediate times these two functions are totally free. 
In particular interpolating $G(t)=\sum_{i=0}^3g_it^i$ and $B(t)=\sum_{i=0}^3b_it^i $ by polynomials with at least the same
degree as the number of boundary conditions lets us deduce from Eq. (\ref{DelOm}) the desired controls $\Delta(t)$ and $\Omega(t)$. However, extra-coefficients can be added to the interpolation, for example, $G(t)=\sum_{i=0}^4g_it^i$. Here $g_4$ can be used to also control the values of the measures $\mathcal{O}$ or $\mathcal{A}$. 
\subsection{Single noise source}
We first consider amplitude noise of the form $\Xopb_1\equiv \sop_z$ and $\eta_1\equiv\eta_z$. 
For this particular type of noise the measure $\mathcal{O}$ is given explicitly by
\beq
\label{Oz}
\mathcal{O}_z=\frac{1}{t_f}\int_0^{t_f} 2\left( \left| \sin\left(\frac{G}{2}\right)\right| +\left|\cos\left(\frac{G}{2}\right)\right|-1\right)dt,
\eeq
which is independent of the free function $B$ of the invariant.
It takes its minimal value $\mathcal{O}_z \rightarrow 0$ when $G(t) \rightarrow  n\pi $ with $n\in \mathbb{Z}$ and the maximal $2\sqrt{2}-2$
when $G(t)\rightarrow \pi/2 + n\pi$. 
Similarly we can write explicitly (\ref{eq:commu_norm}) after some simple algebra:
\beq
\mathcal{A}_z=\frac{\int_0^{t_f}dt \vert \sin(G) \vert}{t_f}.
\eeq
As for the measure $\mathcal{O}_z$ the minimal value $\mathcal{A}_z\rightarrow 0$ occurs when $G\rightarrow n\pi$ and $\mathcal{A}_z\rightarrow 1$ for $G \rightarrow \pi/2 +n\pi$.

\begin{figure}

\begin{minipage}{7.5cm}
\centering
\subfloat[]{\label{fig1:a}
\includegraphics[width=7.5cm]{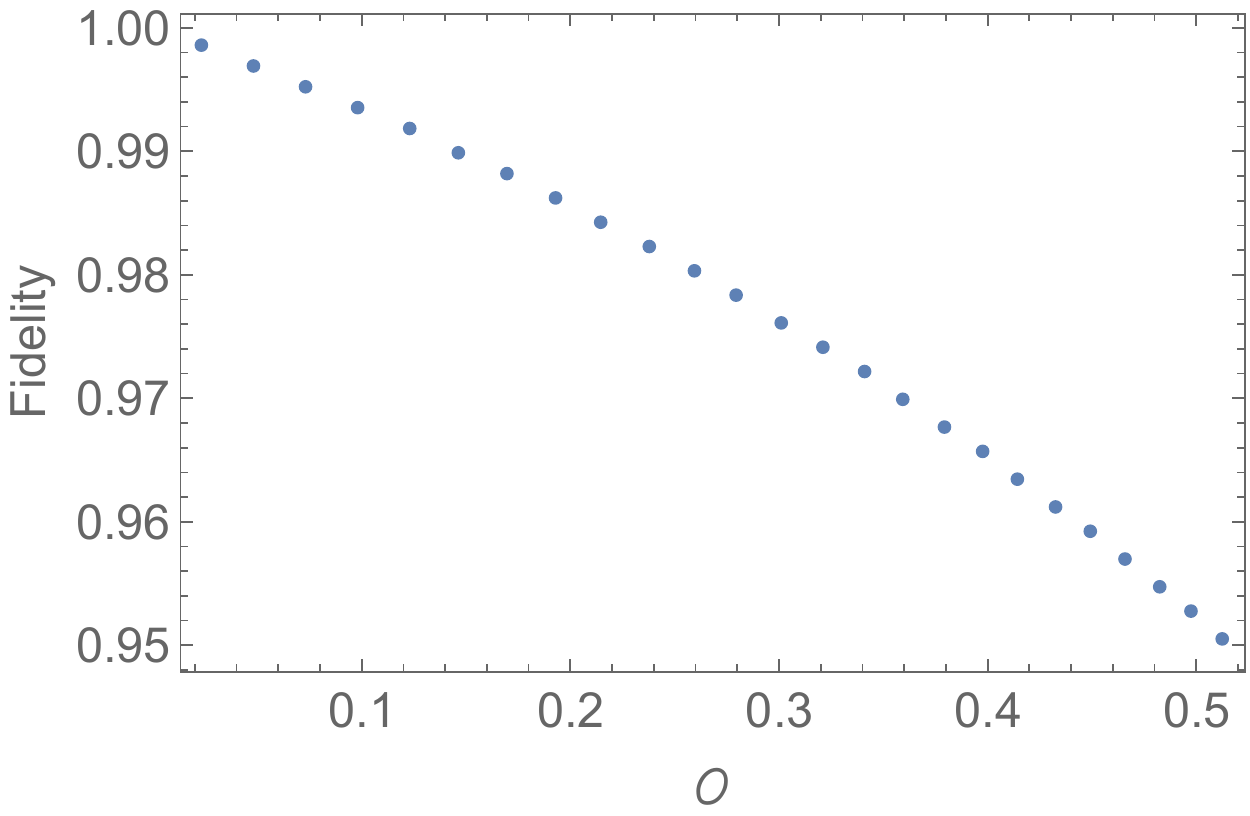}}
\end{minipage}
\qquad
\begin{minipage}{7.5cm}
\centering
\subfloat[]{\label{fig1:b}
\includegraphics[width=7.5cm]{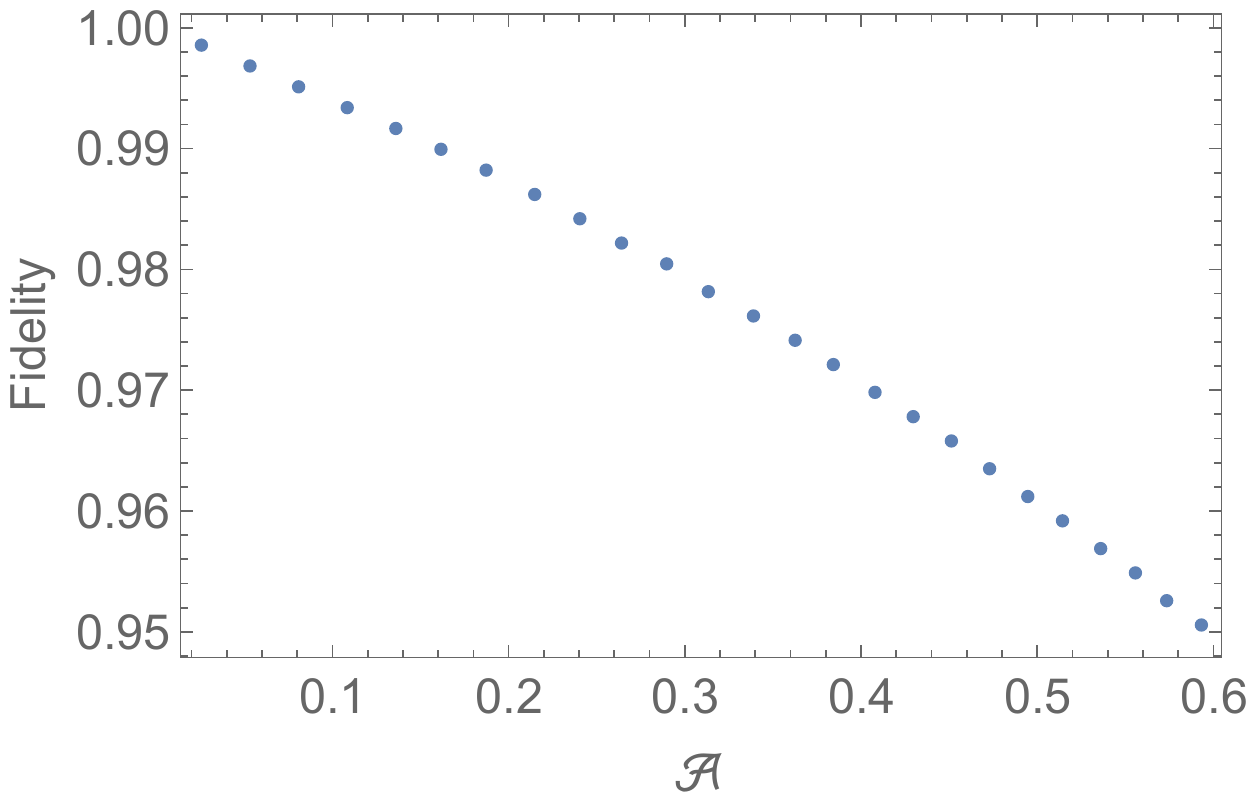}}
\end{minipage}
\caption{ Fidelity as function of the measure $\mathcal{O}$ (a) and the measure $\mathcal{A}$ (b) for a given final time $t_f$.   
Here: $\Delta_0=10$ kHz, $\eta_z=0.25$ kHz and $t_f=0.5$ ms.}
\label{fig:fig1}
\end{figure}

In Figs. \ref{fig:fig1} we plot the fidelity as function of the measures $\mathcal{O}_z$ and $\mathcal{A}_z$ for a given final time $t_f$ for the full population inversion problem.
Both measures show a similar behavior, when
$\mathcal{O}_z \rightarrow 0$ and  $\mathcal{A}_z \rightarrow 0$  the fidelity is improved significantly and monotonically decreasing as $\mathcal{O}_z$ and $\mathcal{A}_z$ increase. 
Thus, by adding constraints on these measures when constructing the invariant we obtain a control field  
which minimizes the effect of the noise.
When the dynamics is subject to noise from a single source, i.e., a single $\Xopb_1$, a control protocol which leads to fidelity $ \simeq 1$ can be found. Generally, when the dynamics is subject to several independent sources of noise,  $\Xopb_j$, obtaining fidelity $ \simeq 1$ is not guaranteed.
Nevertheless, the influence of the overall noise can still be minimized and the final fidelity is improved. 
%
%
%
%
%
\subsection{Multiple noise sources}
\begin{figure}

\begin{minipage}{7.5cm}
\centering
\subfloat[]{\label{fig2:a}\includegraphics[width=8cm]{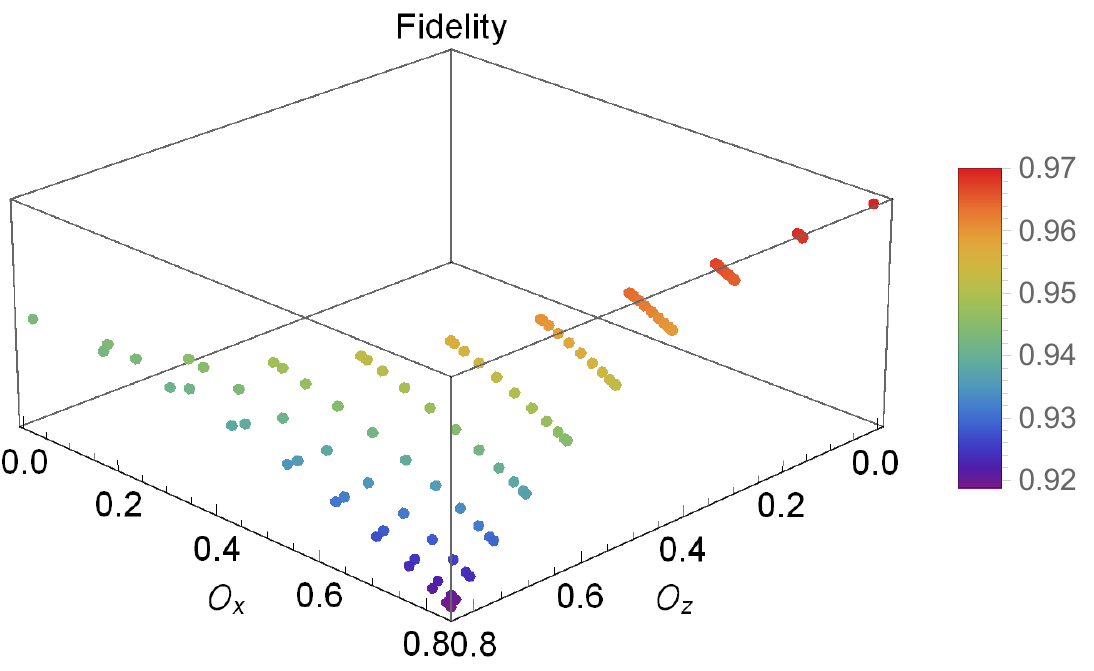}}
\end{minipage}%
\qquad
\begin{minipage}{7.5cm}
\centering
\subfloat[]{\label{fig2:b}\includegraphics[width=8cm]{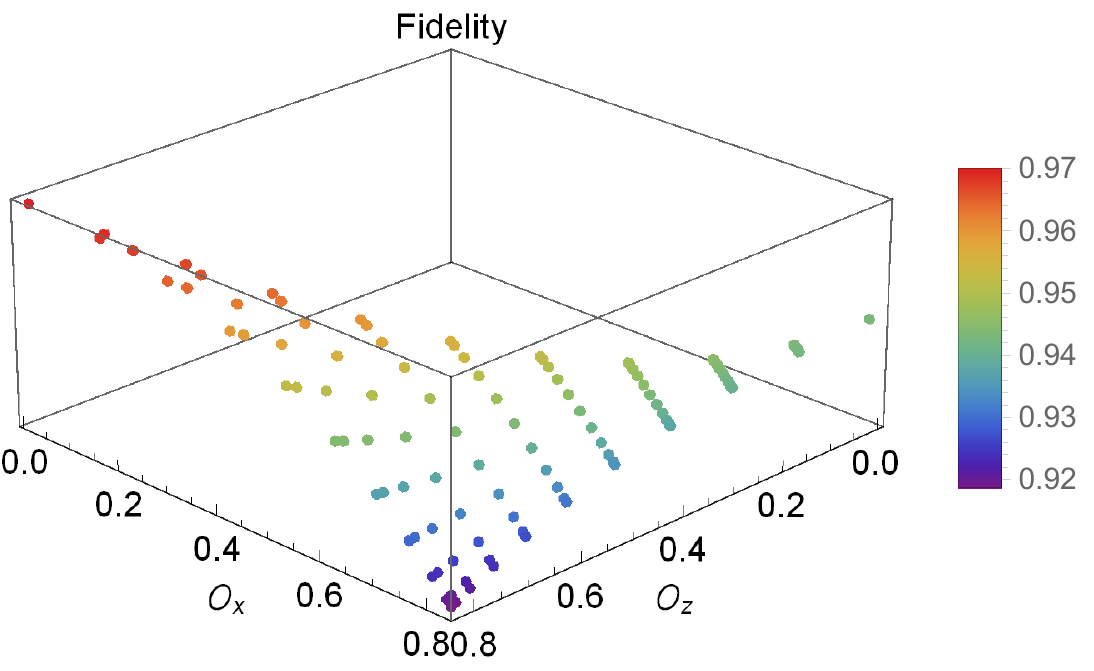}}
\end{minipage} 
\centering
\subfloat[]{\label{fig2:c}\includegraphics[width=8cm]{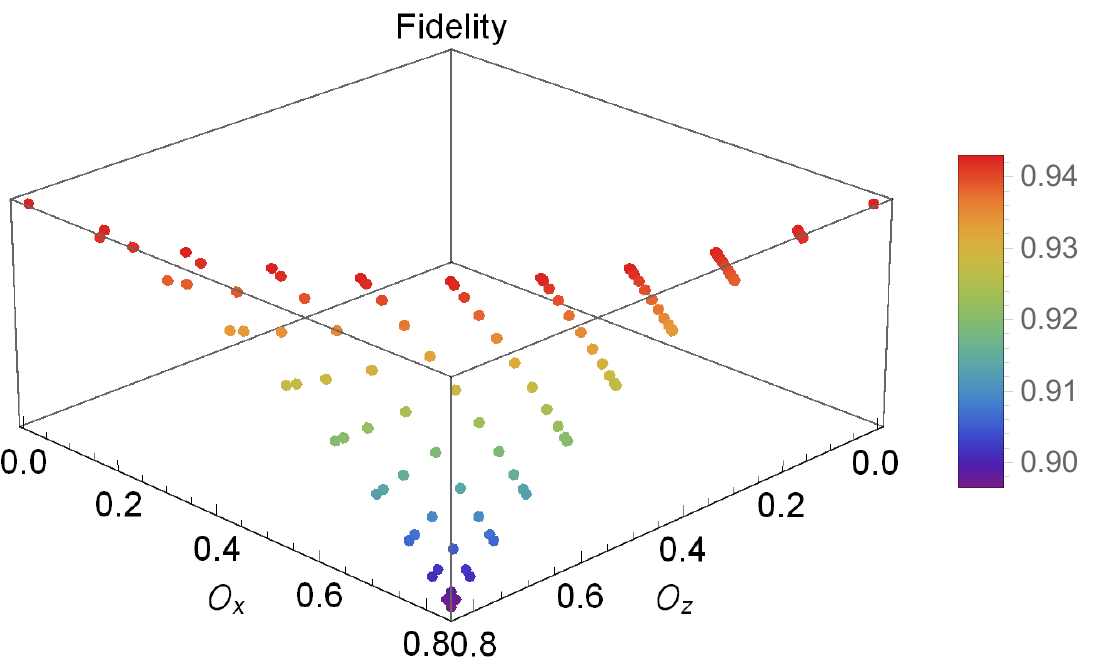}}
\caption{The fidelity of the target state vs. $\mathcal{O}_z$ and $\mathcal{O}_x$ for three different noise strength ratios. (a) $\eta_x=0.125$ kHz and $\eta_z = \eta_x /2$.  (b) $\eta_z=0.125$ kHz and $\eta_x = \eta_z /2$. (c) $\eta_z=\eta_x=0.125$ kHz. In all the plots $\Delta_o=10$ kHz and $t_f=0.5$ ms.
Maximal fidelity is always obtained for minimal $\bar{\mathcal{O}}$.  }
\label{fig:fig2}
\end{figure}

When multiple noise terms (different $\Xopb_j$) are present in the dynamics, the measures $\mathcal{O}_j$ or $\mathcal{A}_j$ cannot always be minimized simultaneously. In the next example we study the worst case scenario where the two noise terms have mutually unbiased bases \cite{durt08}. In this case minimization of one of the noise terms will lead to maximization of the other. 
In particular we consider amplitude noise both in the $\sop_z$ and $\sop_x$ fields, i.e., $\Xopb_1=\sop_z$, $\eta_1=\eta_z$ and $\Xopb_2=\sop_x$, $\eta_2=\eta_x$.
For the TLS we employ $\mathcal{O}_{z(x)}$ to quantify the effect of the noise in the dynamics. As for $\mathcal{O}_z$ we can write explicitly $\mathcal{O}_x$ as 
\beq
\label{Ox}
\mathcal{O}_x=\frac{1}{t_f}\int_0^{t_f} 2\left( \sqrt{\frac{1-\cos(B)\sin(G)}{2}}+\sqrt{\frac{1+\cos(B)\sin(G)}{2}} -1\right)dt. 
\eeq
By examining the integrands of Eqs. (\ref{Oz}) and (\ref{Ox}) we observe that $(i)$ when $G(t) \rightarrow  n\pi $, then $\mathcal{O}_z \rightarrow 0$ and $\mathcal{O}_x$ approach its maximal value $2\sqrt{2}-2$, independently of $B(t)$.  
The other extreme limit is obtained $(ii)$ when $G(t)\rightarrow \pi/2 + n\pi$ and $B(t)\rightarrow n\pi $, then, $\mathcal{O}_z \rightarrow 2\sqrt{2}-2$ and $\mathcal{O}_x \rightarrow 0$.
For multiple noise terms we suggest to minimize the  average $\bar{\mathcal{O}}$ of the single noise measures  weighted according to their relative strength. In the example above this average reads,
\beq
\label{eq:average_O}
\bar{\mathcal{O}}=\frac{\eta_z}{\eta_z+\eta_x}\mathcal{O}_z + \frac{\eta_x}{\eta_z+\eta_x}\mathcal{O}_x,
\eeq 
with a minimum value (see \ref{app:overlap} for more details)
\beq
\label{eq:O_ave}
\bar{\mathcal{O}} \rightarrow (2\sqrt{2}-2) \cdot \min \left\lbrace \frac{\eta_z}{\eta_z+\eta_x},\frac{\eta_x}{\eta_z+\eta_x} \right\rbrace.
\eeq
This implies that in order to optimize the fidelity, protocol $(i)$ or $(ii)$ are chosen depending on the relation of the noise strength $\eta_z$ and $\eta_x$. 
Thus, minimizing the influence of the stronger noise term will lead to higher fidelity as is demonstrated in Fig. \ref{fig:fig2}.
In this figure we plot the fidelity against $\mathcal{O}_z$ and $\mathcal{O}_x$ for different $\eta_z$ and $\eta_x$ ratios.
Maximal fidelity is obtained when the average $\bar{\mathcal{O}}$ is minimal and given by Eq.(\ref{eq:O_ave}).  
We remark that equivalently, optimization can be performed using the measure $\mathcal{A}$ by the replacement  of $\mathcal{O}\rightarrow \mathcal{A}$ in Eq.(\ref{eq:average_O}).

\section{Quantum Harmonic Oscillator}
\label{sec:ho}
In this section we study  the quantum harmonic oscillator which for example  can describe a particle with reduced mass $m$ 
(for the simulations the mass of 100 ions of $^{40}\mbox{Ca}^+$ is used) in a harmonic trap with time dependent frequency $\omega(t)$.
The Hamiltonian takes the form, 
\beq
\label{eq:HO_hamiltonian}
\Hop(t)=\frac{1}{2m}\pop^2 + \frac{m\omega^2(t)}{2}\qop^2.
\eeq
Note that this problem can be mapped to a general control problem of the SU(1,1) algebra \cite{Chen2010, Torrontegui2011, Lewis1969, Diaz2012, Yuste2013}.
Thus, Eq.(\ref{eq:HO_hamiltonian}) can be written as,
\beq
\Hop(t)=a\Top_1 + b(t)\Top_2,
\eeq 
where we define  $a=1/m$, $b(t)=m\omega^2(t)$, and identify $\Top_1=\pop^2/2$, $\Top_2=\qop^2/2$ and $\Top_3=(\pop\qop +\qop\pop)/2$ that satisfy the commutation relations
\beq
[\hat T_1, \hat T_2]=-i \hat T_3,\quad [\hat T_1, \hat T_3]=-2i \hat T_1,\quad [\hat T_2, \hat T_3]=2i \hat T_2.
\eeq
Associated with the harmonic oscillator Hamiltonian (\ref{eq:HO_hamiltonian}) there is a dynamical invariant of the form (see \ref{app:oscillator})
\beq
\hat I(t)=\hat\pi^2/(2m)+m\omega_0^2\hat x^2/2,
\eeq
where $[\hat x,\hat\pi]=i$ with $\hat x\equiv\hat q/\rho$, $\hat\pi\equiv\rho\hat p-m\dot\rho\hat q$, and $\rho$ is an auxiliary scaling function satisfying Ermakov's equation
\beq
\label{rhoeq}
\ddot \rho+\omega^2(t)\rho=\frac{\omega_0^2}{\rho^3},
\eeq
with $\rho(0)=1, \rho(t_f)=\sqrt{\omega_0/\omega_f}$ and $\dot\rho(t_b)=\ddot\rho(t_b)=0$ \cite{Chen2010} imposed by the frictionless conditions $[\hat H(t_b),\hat I(t_b)]=0$ and continuity. As in the example of the TLS we use the freedom
to interpolate the free function $\rho$ at intermediate times. 
We choose functions of polynomials with sufficient parameters to satisfy the previous six boundary conditions. As we showed in the previous section, extra coefficients can be
incorporated with higher order polynomials to impose other constraints such as the minimization of $\mathcal{O}$ or $\mathcal{A}$.

In the next two examples, we study the expansion control of coherent and thermal states. In these cases the success of the control protocol is evaluated according to the previous fidelity definition, Eq. (\ref{fid_def}), for Gaussian states \cite{banchi15}.
\subsection{Coherent states}
We assume that the initial coherent state $\ket\alpha$ with the initial frequency $\omega_0=\omega(0)$ is driven to
the final target state $\ket{\tilde{\alpha}}$ 
with $\omega_f=\omega(t_f)$, where $\tilde\alpha=\alpha e^{-ig\omega_0}$ and $g=\int_0^{t_f}dt'/\rho^2$.
For this end we interpolate $\rho(t)=\big( \sum_{i=0}^5r_it^i\big)^{-1/2}$ and deduce $\omega(t)$ from Eq. (\ref{rhoeq}) (see \ref{app:oscillator}).   
This noise arises from weakly and continuously measuring (monitoring) the position of the particle in the trap leading to
$\Xopb=\qop$ \cite{wisemanbook, diosi88}. 
As was discussed above, for unbounded operators the calculation of the overlap between the bases to compute $\mathcal{O}$ should be carried on a finite domain or as we will see next it can be evaluated using
\beq
\label{eq:HO_measure}
S_n=\frac{1}{t_f}\int^{t_f}_0\int^{\infty}_{-\infty}\vert\braket{\qop}{\phi_n(t)}\vert dqdt.
\eeq
This overlap can be written explicitly as (see \ref{app:oscillator})
\beq
\vert \langle q|\phi_n(t)\rangle \vert =\frac{ \sqrt[4]{\frac{ m\omega_0} { \pi} }e^{-m\omega_0 q^2/2\rho^2}}{\sqrt{2^n n!}\sqrt{\rho}} \vert H_n(\sqrt{m\omega_0}q/\rho) \vert,
\eeq
where $H_n$ are the Hermite polynomials.
In principle, to compute $\mathcal {O}$ we should consider the sum over $n$ from $0$ to $\infty$ of the elements $S_n$ in Eq. (\ref{eq:HO_measure}).
Nevertheless, we find that minimizing Eq. (\ref{eq:HO_measure}) for a certain $n$ will necessarily  minimizes all the different $n$ terms.
We prove this by showing that the spatial integration over $q$ is independent of the function $\rho$.
\par
Proof: We preform the following coordinate substitution, $u_1=q/\rho$ and $u_2=1/\sqrt{\rho}$. The determinant of the Jacobian is given by
\beq
\det \left[\begin{array}{cc}
  \frac{\partial u_1}{\partial q}       &   \frac{\partial u_1}{\partial t}      \\
  \frac{\partial u_2}{\partial q}       &   \frac{\partial u_2}{\partial t}     
\end{array} \right] =
\det \left[\begin{array}{cc}
  \frac{1}{\rho}       &   -\frac{q\dot{\rho}}{\rho^2}      \\
  0      &   -\frac{\dot{\rho}}{2\rho^{3/2}}     
\end{array} \right]
= \dot{u}_2 u_2^2.
\eeq
Then, Eq. (\ref{eq:HO_measure}) takes the form 
\beq
\frac{\sqrt[4]{\frac{ m\omega_0} { \pi} }}{t_f \sqrt{2^n n!}}\int^{u_2(t_f)}_{u_2(0)}\vert\dot{u}_2\vert u_2^3du_2 \int^{\infty}_{-\infty} 
e^{-m\omega_0 u_1^2/2}\vert H_n(\sqrt{m\omega_1} u_1)\vert du_1.
\eeq 
The integration over $u_1$ depends on $n$, but it is independent of $\rho$. 
Thus, different designs of $\rho$ influence only the integration over $u_2$ which is independent of $n$, implying that it is sufficient to minimize Eq. (\ref{eq:HO_measure}) for an arbitrary $n$ when constructing the invariant.
\par
\begin{figure}
\begin{minipage}{7.5cm}
\centering
\subfloat[]{\label{fig6:a}
\includegraphics[width=7.5cm]{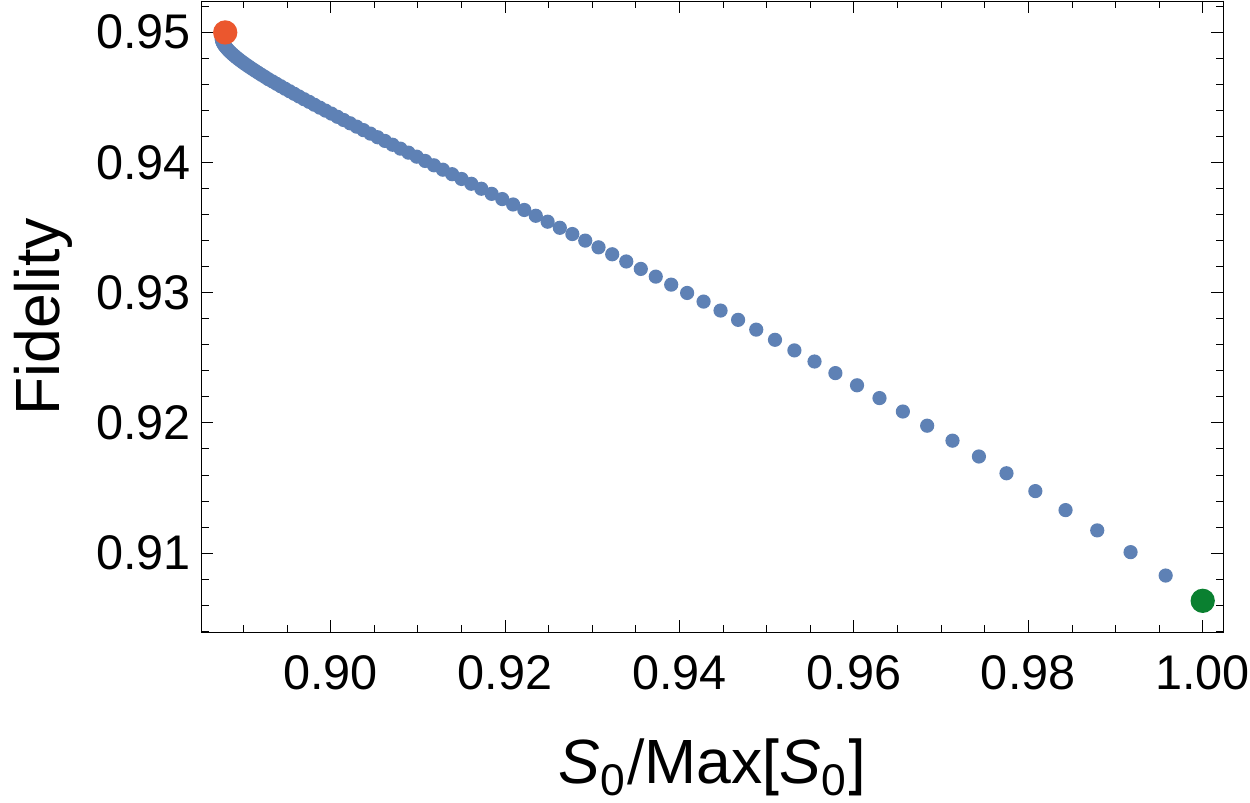}}
\end{minipage}
\qquad
\begin{minipage}{7.5cm}
\centering
\subfloat[]{\label{fig6:b}
\includegraphics[width=7.5cm]{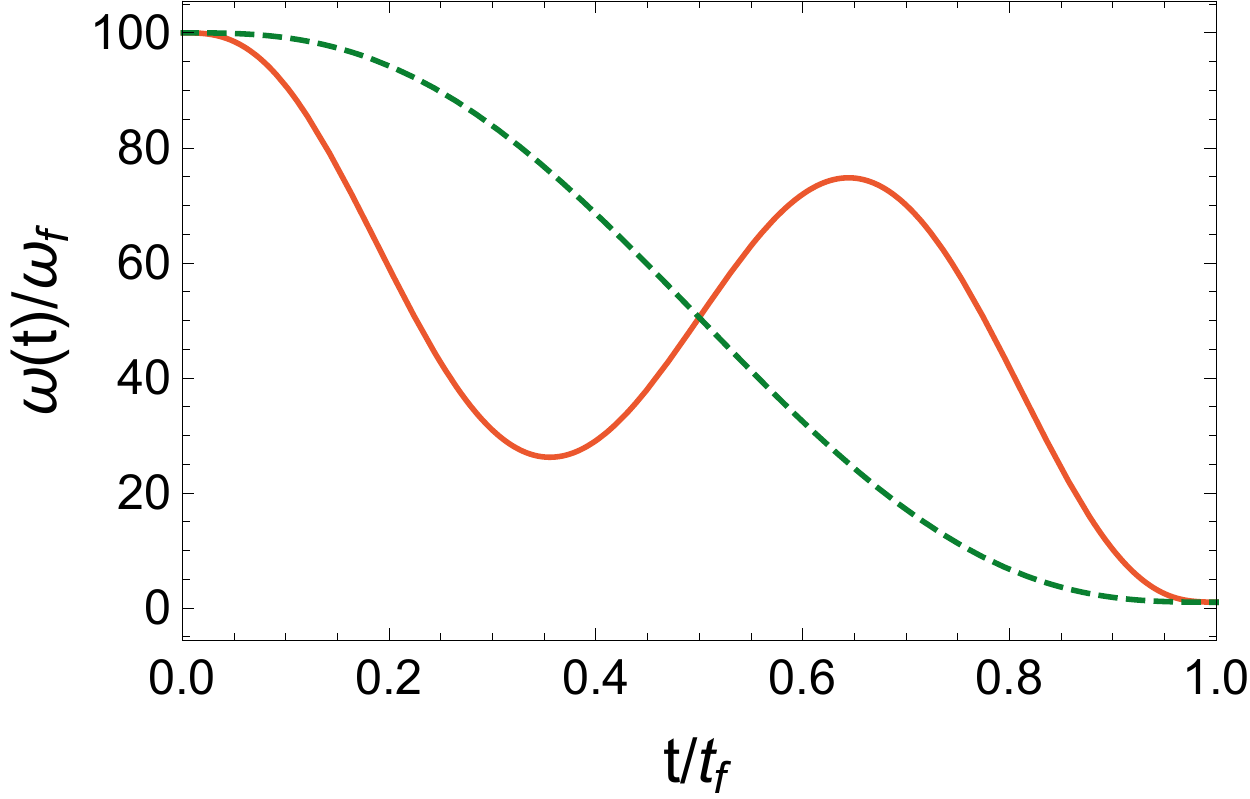}}
\end{minipage}
\caption{ (a) Fidelity vs. the normalized $S_0$. Each point in the figure corresponds to a different control. (b) The control frequency $\omega(t)$ as a function of $t$ for the green (dashed) and red (solid) points. 
Parameter values: $\nu_0=\omega_0/(2\pi)= 15.92$ MHz, $\omega_f=\omega_0/100$,  $\eta=10$ Hz$\AA^{-2}$, and $t_f=100$ $\mu$s. The initial coherent state is given by  $\alpha =1+i$ and the final state by $g=50.5$ $\mu$s. 
}
\label{fig:fig3}
\end{figure}

In Fig. \ref{fig6:a} we design different protocols and plot the fidelity against $S_0$ normalized by the maximal $S_0$ value out of the protocols 
considered in the figure. This is done by
adding two extra-coefficients in the invariant interpolation $\rho(t)=\big(\sum_{i=0}^7r_it^i\big)^{-1/2}$, where $r_6$ and $r_7$ control the values of $S_0$ in Eq. (\ref{eq:HO_measure}) and $g$ that let us fix the final target coherent state independently of the $\rho$ interpolation (see Eq. (\ref{coh_state}).)
As $S_0$  becomes smaller the fidelity is enhanced. 
Figure \ref{fig6:b} presents two control protocols corresponding to the green and red points of Fig. \ref{fig6:a}. 
The green dashed line represents the standard STA protocol \cite{Chen2010} (standard refers to those protocols where the free functions in the invariant are only constrained by the boundary frictionless conditions). 
This protocol can be improved using the method we presented, minimizing $S_0$ to achieve higher fidelities.
We remark that higher fidelity than those shown in Fig. \ref{fig:fig3}a can be achieved just if a higher order polynomial is incorporated when interpolating $\rho$.
%
%
%
%
%
%
%
%
%
%
%
%
%
%
\subsection{Thermal states} 
Consider again the harmonic oscillator Hamiltonian (\ref{eq:HO_hamiltonian}), we now choose different states to protect against noise.
The initial state is assumed to be  the thermal state, $\rop_{0} = \exp(-\beta_0 \Hop(0))/Z $, with the normalization factor $Z$, and the initial inverse temperature
$\beta_0$ and frequency $\omega_0 \equiv \omega(0)$.
The final Hamiltonian corresponds to the frequency $\omega_f \equiv \omega(t_f)$ and the target state is the thermal state $\rop_{tar} = \exp(-\beta_f \Hop(t_f))/Z' $ with the final inverse temperature $\beta_f=\beta_0 \omega_0/\omega_f$.
The noise considered in this example is  noise in the modulation of the frequency described by the noise operator $\Xopb =2\Top_2=\hat q^2$ and constant $\eta$. 
Since this noise is more problematic for long operation times, a natural way to avoid it is to have short operation times.
However, very short expansion times are typically not feasible experimentally.
Designing protocols protected against amplitude noise improve the final fidelities even at longer times.
\par
\begin{figure}
\begin{minipage}{7.5cm}
\centering
\subfloat[]{\label{fig4:a}
\includegraphics[width=7.5cm]{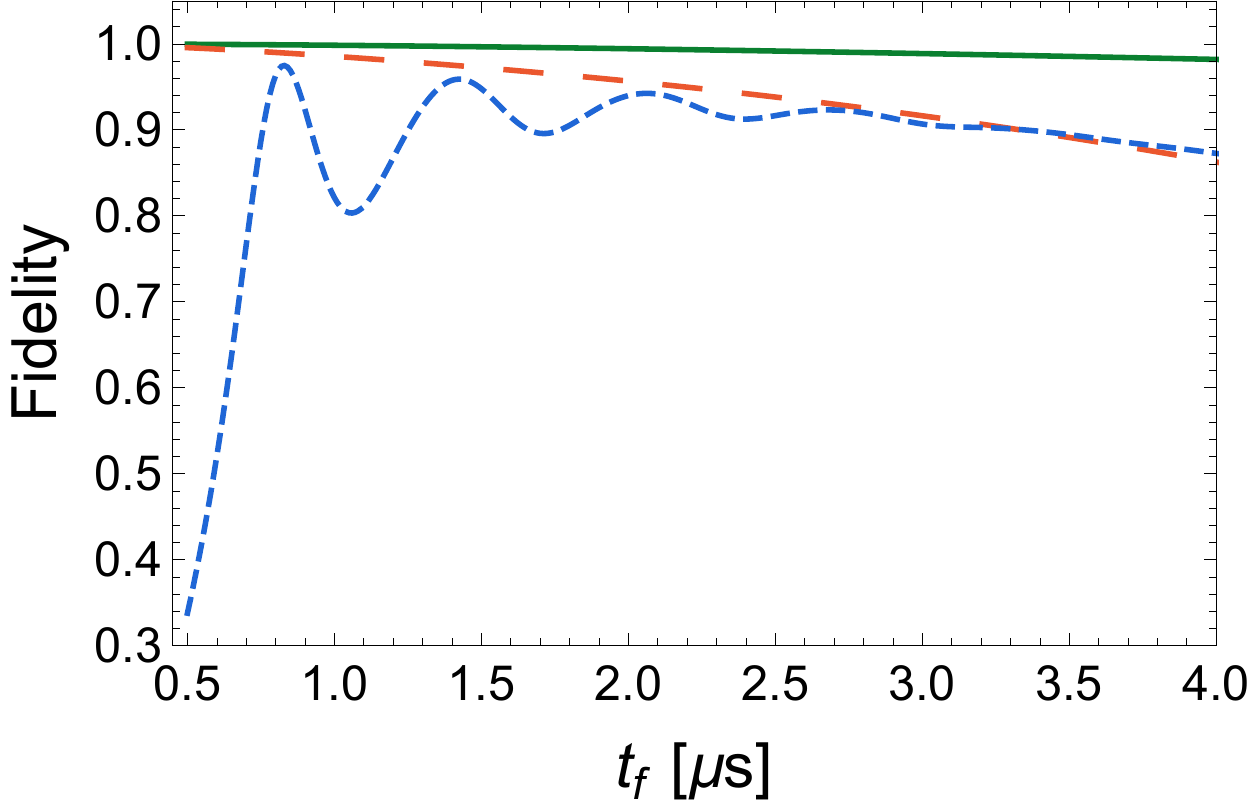}}
\end{minipage}
\qquad
\begin{minipage}{7.5cm}
\centering
\subfloat[]{\label{fig4:b}
\includegraphics[width=7.5cm]{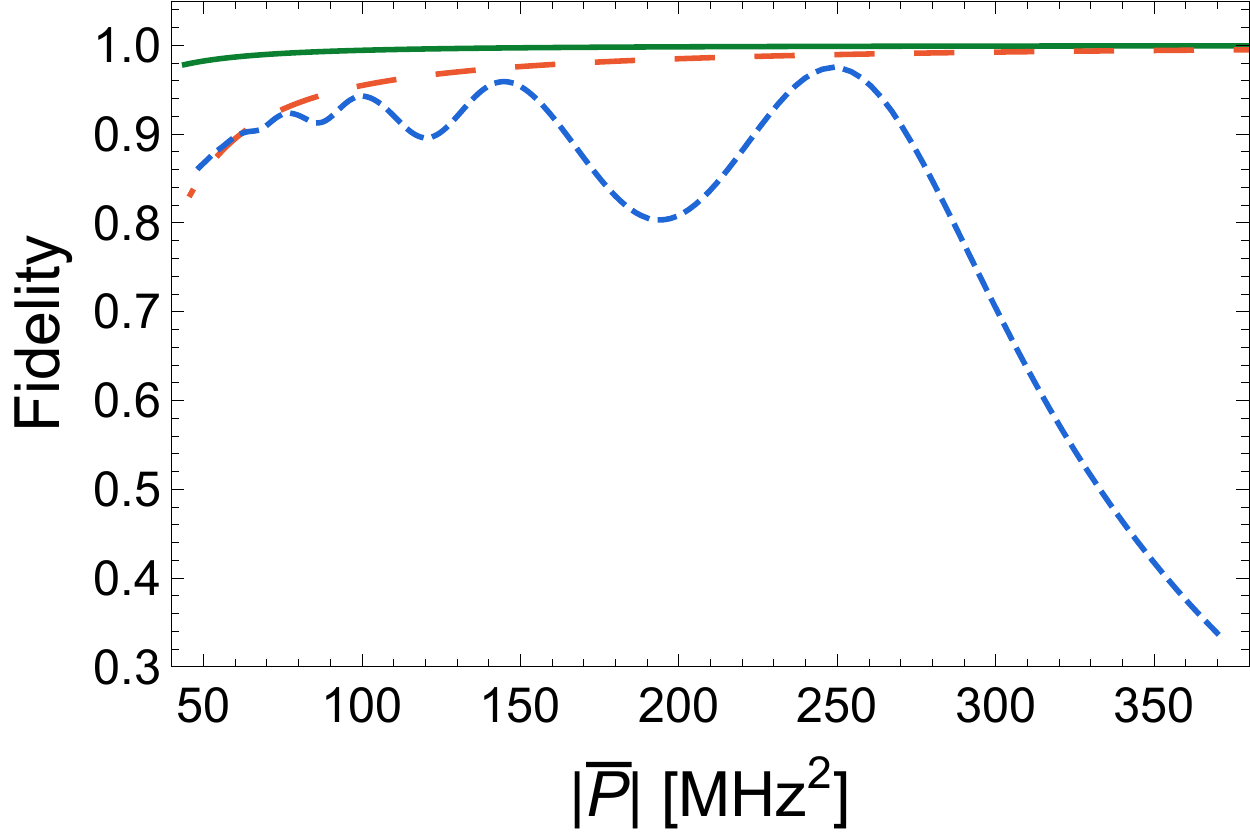}}
\end{minipage}
\caption{ Fidelity as function of (a) the final time $t_f$ and (b) the power $\bar{\mathcal{P}}$, absolute value of the integral (\ref{power}).
In blue (short-dashed) the fast adiabatic protocol of constant $\mu$, in red (long-dashed) the standard STA protocol and in green (solid) the improved STA protocol.
Parameter values: $\nu_0=\omega_0/(2\pi)=2.53$ MHz, $\omega_f=\omega_0/100$, and $\eta=0.0527$ Hz$\AA^{-4}$. The initial state has a temperature of
$T_0=10$ mK and an average occupation number $\bar n=12.58$. }
\label{fig:fig4}
\end{figure}

In Fig. \ref{fig:fig4}a we plot the fidelity against final times $t_f$ for three different control protocols. In blue we plot the fast adiabatic protocol of constant $\mu \equiv \dot{\omega}/\omega^2$  \cite{levy17}, in red the standard STA protocol, and in green the improved STA protocol 
(for both STA protocols $\omega(t)$ is deduced from Eq. (\ref{rhoeq}) using the following ansatzes: $\rho(t)=\big(\sum_{i=0}^5 r_it^i\big)^{1/2}$ for the standard and $\rho(t)=\big(\sum_{i=0}^6 r_it^i\big)^{1/2}$ for the improved protocols, respectively). 
We see that for the optimized STA protocol higher fidelity for all final times $t_f$ is obtained. 
This introduces high flexibility for controlling the final time and the average instantaneous power consumption/production,
\beq
\label{power}
\bar{\mathcal{P}} \defeq \frac{1}{t_f}\int_0^{t_f} \ave{\frac{\partial \Hop(t)}{\partial t}}dt.
\eeq 
In Fig. \ref{fig:fig4}b we plot the fidelity vs. the absolute value of the averaged instantaneous power $\bar{\mathcal{P}}$.
\section{Discussion}
\label{sec:discussion}
In this work we introduce a  method to construct a control protocol  which is robust against dissipation of the population and minimizes the effect of dephasing. Doing so, we optimize the fidelity of the final state with respect to the target state.
As is shown in Eq.(\ref{eq:rho_limits}) the diagonal terms of the density matrix will remain constant at the end of the process while the off diagonal will be affected by dephasing at a rate proportional to the square of the distance between the eigenvalues of the noise operator.
This is a clear indication that Markovian noise cannot be completely suppressed without adding an auxiliary system which will store the information about the coherence.
\par     
The idea of the method presented is based on the fact that the dynamical invariant provides a family of infinite solutions from which the control Hamiltonian can be constructed for a particular state transfer problem.       
By imposing additional constraints on the invariant, namely, minimization of the measure $\mathcal{O}$ or equivalently $\mathcal{A}$, we protect the invariant from the noise during the process. 
Since at the final time the invariant and the Hamiltonian share common eigenvalues the target state is achieved with high fidelity.
The main advantages of this method compared to other control optimization methods is its simple implementation. 
It does not require calculations by iteration and does not involve perturbation methods.
Since the structure of the invariant for the unitary dynamics is already known in many cases, imposing additional constraints on this invariant is not a difficult task. 
Moreover, the method is applicable to different time scales, from the nonadiabatic to the adiabatic regime. 
This implies that in order to suppress the noise we are not limited to frequent sudden operations which in many cases are not feasible experimentally and will typically be costly in terms of power. 
Formulation of the method in terms of density operator is necessary to treat noise but it also makes controlling mixed states possible as in the example of the thermal state.
\par   
The idea of protecting the invariant from the noise during the process and by that designing an optimal noise resistant control can in principle be applied to all types of noise including thermal and non-Markovian noise. 
If the noise operators are not  Hermitian,
we suggest (like in the procedure above) to find the invariant for the unitary dynamics with additional degrees of freedom which can be set later to minimize the effect of noise. 
Next, instead of considering the overlap between the bases which now might not be computable, we can use the rate of change of the eigenvalues subject to the full dynamics as a measure for minimization.
In the limit $\sum_l \vert\dot{\lambda}_l\vert \rightarrow 0$ the noise will not affect the invariant and a noise resistant control can be found.

\ack
We acknowledge L. McCaslin for fruitful discussions,  
funding by the Israeli Science Foundation, the US Army Research Office under Contract W911NF- 15-1-0250,
the Basque Government (Grant No. IT986-16), 
MINECO/FEDER,UE (Grants No. FIS2015-70856-P and No. FIS2015-67161-P), and QUITEMAD+CM S2013-ICE2801. \\
\appendix
\section{Invariant inverse engineering based on Lie algebras}
\label{app:general}
We summarize \cite{Torrontegui2014}, a systematic approach to inverse engineering the controls from the dynamical invariants of a system
when it is described by a closed Lie algebra.
Let us assume that the time-dependent Hamiltonian $\hat H(t)$ describing a quantum system  
is given by a linear combination of Hermitian generators $\hat T_a$,
\beq
\label{ham}
\hat  H(t)=\sum_{a=1}^{N}h_a(t) \hat T_a,
\eeq
where the $h_a(t)$ are real time-dependent functions 
and the $\hat  T_a$ span a  
Lie algebra \cite{Kaushal1981}
\beq
\label{lie}
[\hat  T_b, \hat  T_c]=\sum_{a=1}^N\alpha_{abc} \hat  T_a,  
\eeq
with $\alpha_{abc}$ the structure constants. Associated with the Hamiltonian
there are time-dependent Hermitian invariants of motion $\hat  I(t)$ that satisfy \cite{lewis69}
\beq
\label{def}
\frac{d \hat I}{dt}\equiv\frac{\partial \hat I(t)}{\partial t}-\frac{1}{i}[\hat  H(t), \hat  I(t)]=0.
\eeq
A wave function $|\Psi(t)\rangle$ which evolves 
with $\hat  H(t)$ 
can be expressed as a linear superposition of the instantaneous invariant modes  \cite{lewis69}
\beq
\label{wave}
|\Psi(t)\rangle=\sum_{n}c_ne^{i\alpha_n}|\phi_n(t)\rangle,
\eeq
where the $c_n$ are constants, the phases $\alpha_n$ fulfill
\beq
\label{appeq:phase}
\frac{d\alpha_n}{dt}=\langle\phi_n(t)|i\frac{\partial}{\partial t}-\hat  H(t)|\phi_n(t)\rangle,
\eeq
and the eigenvectors $|\phi_n(t)\rangle$ of $\hat  I(t)$ 
\beq
\hat  I(t)|\phi_n(t)\rangle=\lambda_n|\phi_n(t)\rangle,
\eeq
where $\lambda_n$ are the constant eigenvalues. 

If the invariant is also a member of the dynamical algebra, it can be written as
\beq
\label{inv}
\hat  I(t)=\sum_{a=1}^{N}f_a(t) \hat  T_a,
\eeq
where $f_a(t)$ are real, time-dependent functions.
Replacing  Eqs. (\ref{ham}) and (\ref{inv})
into Eq. (\ref{def}), and using Eq. (\ref{lie}), the  functions $h_a(t)$ and $f_a(t)$ satisfy \cite{Kaushal1981,Maamache1995}
\beqar
\dot f_a(t)&=&\sum_{b=1}^{N}\mathcal{G}_{ab}(t)h_b(t), \nonumber
\\
\label{compact}
&{\rm or}&  |\mathbf{\dot f}\rangle=\mathcal{G}|\mathbf{h}\rangle, 
\eeqar
with the $N\times N$ matrix $\mathcal{G}$ 
\beq
\label{A}
\mathcal{G}_{ab}(t)=\frac{1}{i}\sum_{c=1}^{N}\alpha_{abc}f_c(t),
\eeq
where the kets are defined in terms of the component of each generator \cite{Torrontegui2014}. Note 
that the relation between the Hamiltonian and the invariant is a property of the algebra, i.e. the structure constants, and is independent of the representation.

Usually these coupled equations are interpreted as a linear system of ordinary differential 
equations for $f_a(t)$ when the $h_a(t)$ components of the Hamiltonian are known \cite{Kaushal1981,Maamache1995,Kaushal1993,Monteoliva1994}.
Here we consider a different perspective taking them as an algebraic system to be solved for 
the $h_a(t)$, when the $f_a(t)$ are given. 
As there are  many Hamiltonians for a given invariant \cite{Chen2011}   
we cannot generally invert Eq. (\ref{compact}) as $|\mathbf{h}\rangle=\mathcal{G}^{-1}|\mathbf{\dot f}\rangle$ to get $|\mathbf{h} \rangle$.  
This means that det($\mathcal{G}$)$=0$, so at least one of the eigenvalues  $a^{(i)}(t)$ of the $\mathcal{G}$ matrix vanishes.  
Different approaches, such as Gauss elimination or projector techniques \cite{Torrontegui2014},  can be used to find the pseudo-inverse matrix of $\mathcal{G}$ and deduce the 
Hamiltonian component $|\mathbf{h}\rangle$ in terms of the invariant $|\mathbf{f}\rangle$.

When inverse engineering shortcuts to adiabaticity \cite{Chen2010, torrontegui13}, the Hamiltonian is usually given at initial and final times. 
In general the invariant $\hat I$ (equivalently $|\mathbf{f}(t)\rangle$) is chosen   
to drive, through its eigenvectors,  the initial states of the Hamiltonian $H(0)$ to the states of the final $\hat H(t_f)$ \cite{Chen2010, Torrontegui2011,lewis69} according to Eq. (\ref{wave}). 
This is ensured by imposing at the boundary times $t_b=0, t_f$,  the frictionless conditions $[\hat  H(t_b), \hat  I(t_b)]=0$ \cite{Chen2010}.
Equivalently, using
Eqs. (\ref{ham}), (\ref{inv}), and since the $\hat T_a$ generators are independent this condition implies
\beq
\label{BC}
\mathcal{G}(t_b)|\mathbf{h}(t_b)\rangle=0,\quad t_b=0,t_f. 
\eeq
At the boundary times Eq. (\ref{BC}) imposes $N$ conditions, however, 
at intermediate times the Hamiltonian and invariant components can be freely designed    
subjected to the $N$ equations in Eqs. (\ref{compact}).
This leaves open different inverse engineering 
possibilities: in general the Hamiltonian is first fixed partially, i.e., imposing the time dependence (or vanishing) of 
some $r<N$ components. Fixing the invariant time dependence consistently with the boundary conditions and the imposed 
Hamiltonian constraints, finally leads to equations that give the form of the remaining $N-r$ Hamiltonian components.  
\section{The SU(2) algebra and the two-level system}
\label{app:su2}
Let us consider a system where the commutation relations of the generators span a SU(2) Lie algebra
\beq
[\hat T_1, \hat  T_2]=i \hat  T_3,\; [\hat  T_2, \hat  T_3]=i \hat  T_1,\;[\hat  T_3, \hat  T_1]=i \hat T_2.
\eeq
The relation among the Hamiltonian and invariant components, Eq.(\ref{compact}), becomes
\beq
\label{sisSU2}
\left(\begin{array}{c} 
\dot f_1 \\
\dot f_2   \\
\dot f_3 
\end{array} \right)=
\underbrace{\left(\begin{array}{ccc} 
0&   f_3 & -f_2  \\
-f_3 &   0& f_1 \\
f_2      &  -f_1 & 0 
\end{array} \right)}_{=\mathcal{G}}
\left(\begin{array}{c} 
h_1  \\
h_2\\
h_3
\end{array} \right).
\eeq
As we pointed before for this algebra $\det(\mathcal{G})=0$, so Eq. (\ref{sisSU2}) is not directly invertible. After some
simple algebra we find the $h_a(t)$ components in terms of $f_a(t)$ if the constraint 
\beq
\label{cSU2}
\dot f_1f_1+\dot f_2f_2+\dot f_3f_3=0
\eeq
or equivalently $f_1^2+f_2^2+f_3^2=c$ is fulfilled then,
\beq
\label{Hs1_sol}
h_i=-{\cal{E}}_{ijk}\frac{\dot f_j}{f_k}+\frac{f_i}{f_k}h_k,
\eeq
with all indices $i,\, j,\, k$ different, ${\cal{E}}_{ijk}$ is the Levy-Civita 
symbol (1 for even permutations of (123) and -1 for odd permutations), $c$ is a constant,  
and $h_k(t)$ is considered a Hamiltonian free component chosen 
for convenience. The frictionless conditions (\ref{BC}) for this algebra is
\beq
\label{fricSU2}
f_i(t_b)h_j(t_b)-f_j(t_b)h_i(t_b)=0,\;\;  i>j.
\eeq

To be more specific note that the TLS in Sec. \ref{sec:two_level} is governed by this algebra with the following representation of generators,
\beq
\!\!\!\!\!\!\hat T_1=\frac{1}{2}\left(\begin{array}{cc} 
0 & 1 \\
1 & 0
\end{array} \right)\!,
\hat T_2=\frac{1}{2}\left(\begin{array}{cc} 
0 & -i \\
i & 0
\end{array} \right)\!, 
\hat T_3=\frac{1}{2}\left(\begin{array}{cc} 
1 & 0 \\
0& -1
\end{array} \right)\!, 
\eeq
where $h_1(t)=\Omega(t)$, $h_3(t)=\Delta(t)$, and the boundary Hamiltonians $\Omega(0)=0$, $\Delta(0)=\Delta_0$ at $t=0$ and $\Omega(t_f)=0$, $\Delta(t_f)=-\Delta_0$ at $t=t_f$ to produce
the population inversion among the $|0\rangle$ and $|1\rangle$ states.
The objective is to design $\Omega(t)$ and $\Delta(t)$ to connect these two states by imposing partially the structure Eq. (\ref{H2}) of $\hat H(t)$, i.e. $ h_2(t)=0$ $\forall$t,
as it is not always experimentally feasible to implement $\hat T_2$. Imposing $h_2(t)$ we chose to interpolate $f_1$ and $f_2$ satisfying the boundary conditions $f_1(t_b)=f_2(t_b)=\dot f_1(t_b)=\dot f_2(t_b)=0$
imposed by (\ref{fricSU2}). We use polynomial interpolations with at least the same
degree as the number of boundary conditions, nevertheless, higher order polynomials can be considered to 
impose even more constraints. Then $f_3$ is given by (\ref{cSU2}) with $f_3(t_b)=h_3(t_b)\sqrt{c/[h_1^2(t_b)+h_2^2(t_b)]}$ and $\dot f_3(t_b)=0$.
Once the $f_a$ are fixed $\Omega(t)$ and $\Delta(t)$ are deduced from Eq. (\ref{Hs1_sol}),
\beq
\label{GB_su2}
\Delta=-\frac{\dot f_1}{f_2}, \quad \Omega=\frac{\dot f_3}{f_2}.
\eeq
An alternative and sometimes convenient choice to express the invariant is using the angles on the Bloch sphere $G(t)$ and $B(t)$, parametrazing 
$f_1(t)=\Omega_R\sin (G(t))\cos (B(t))$, $f_2(t)=\Omega_R\sin (G(t))\sin (B(t))$, and choosing $\Omega_R=\sqrt{\Omega^2(0)+\Delta^2(0)}$.
The constraint (\ref{cSU2}) imposes $f_3(t)=\Omega_R\cos (G(t))$ with $c=\Omega_R^2$, so the invariant $\hat I$ is expressed as in Eq. (\ref{Isu2}). According to Eq. (\ref{Hs1_sol}) the Hamiltonian coefficients $h_a(t)$ are given in terms of these polar angles  
as \cite{Chen2011}
\beq
\Delta=-\dot B+\frac{\dot G}{\tan (G)\tan (B)}, \quad \Omega=\frac{\dot G}{\sin (B)},
\eeq
with the boundary conditions for population inversion $G(0)=\pi$, $G(t_f)=0$, $\dot G(t_b)=0$, remaining $B(t_b)$ and $\dot B(t_b)$ as free parameters.
\section{Overlap Matrix for TLS}
\label{app:overlap}
\begin{figure}[t]
\center{\includegraphics[width=8.5cm]{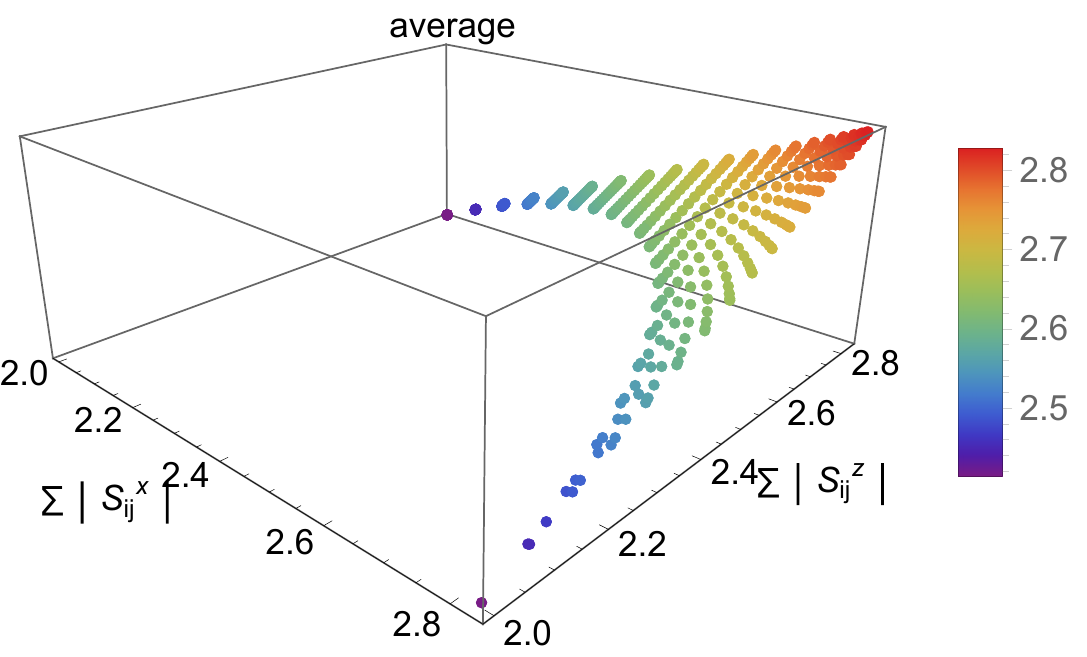}}
\caption{ The weighted average overlap Eq. (\ref{app:eq:ave}) as function of $\sum_{ij}\vert S_{ij}^z \vert$ and $\sum_{ij}\vert S_{ij}^x \vert $ for $p=\frac{1}{2}$.}
\label{fig:fig7}
\end{figure}
For two arbitrary bases in the Hilbert space $\mathcal{C}^2$ the overlap matrix $S$ is bounded by 
\beq
2\leqslant \sum_{ij=1}^2\vert S_{ij} \vert \leqslant 2\sqrt{2}.
\eeq
In this scenario there are three bases that maximize the overlap $\sum_{ij}\vert S_{ij} \vert$. 
These bases are the mutually unbiased bases given by the eigenvectors of the Pauli matrices. 
In the main text we study the simultaneous overlap of two mutually unbiased bases with the basis of the invariant. 
In particular the bases of interest are the eigenvectors of $\sop_z$  and $\sop_x$ which reads 
$\left\lbrace (1,0)^t, (0,1)^t\right\rbrace$ and   $\left\lbrace (1/\sqrt{2},1/\sqrt{2})^t, (1/\sqrt{2},-1/\sqrt{2})^t\right\rbrace$ respectively ($t$ means transpose).
An arbitrary basis in $\mathcal{C}^2$ can be expressed in terms of two real parameters $\theta\in [0,\pi]$ and $\varphi \in [0,2\pi]$,
\beqar
\label{app:eq:basis}
 \left(\sqrt{ 1+\e^{i \varphi}\tan(\theta/2)^2}\right)^{-1} 
\left(\begin{array}{cc} 
 -\e^{i \varphi}\tan(\theta/2) \\
1 
\end{array} \right) ,\\ \nonumber
 \left(\sqrt{ 1+\e^{i \varphi}\cot(\theta/2)^2}\right)^{-1} \left(\begin{array}{cc} 
 -\e^{i \varphi}\cot(\theta/2) \\
1 
\end{array} \right).  
\eeqar 
The weighted average of the overlaps of the two bases with \ref{app:eq:basis} is given by
\beqar
\label{app:eq:ave}
p\sum_{ij}\vert S_{ij}^z \vert + (p-1)\sum_{ij}\vert S_{ij}^x \vert = \\ \nonumber
 2p\left( \left| \sin\left(\frac{\theta}{2}\right)\right| +
 \left|\cos\left(\frac{\theta}{2}\right)\right|\right)+ \\ \nonumber
 2(p-1)\left( \sqrt{\frac{1-\cos(\varphi)\sin(\theta)}{2}}+\sqrt{\frac{1+\cos(\varphi)\sin(\theta)}{2}} \right).
 \eeqar
where the weight $p\in [0,1]$.
Since the two overlaps depends on each another, not all values of the overlaps are reachable simultaneously. This is shown in Fig. \ref{fig:fig7}, where the average (\ref{app:eq:ave}) is plotted  as function of the overlaps  $\sum_{ij}\vert S_{ij}^z \vert$ and $\sum_{ij}\vert S_{ij}^x \vert $ for $p=\frac{1}{2}$.
\par
The minimum average is given by
\beq
\label{app:eq:min _ve}
2\sqrt{2}\min(p,p-1)+2\max(p,p-1),
\eeq
 and is always obtained when one of the overlaps is minimal and the other is maximal (as expressed in \ref{app:eq:min _ve}). 
The maximum average is obtained when the third basis is one of the eigenvectors of $\sop_y$, i.e. $\left\lbrace (i/\sqrt{2},1/\sqrt{2})^t, (i/\sqrt{2},1/\sqrt{2})^t\right\rbrace$, and is given by the maximal overlap $2\sqrt{2}$.
In the special case $p=\half$ two minimal points can be found, as shown in Fig. \ref{fig:fig7}, whether we minimize $\mathcal{O}_x$ and maximize $\mathcal{O}_z$ or vice versa. 
\section{The SU(1,1) algebra and the harmonic oscillator}
\label{app:oscillator}
The SU(1,1) algebra is characterized by the commutation relation
\beq
\label{su11}
[\hat T_1, \hat T_2]=-i \hat T_3,\quad [\hat T_1, \hat T_3]=-2i \hat T_1,\quad [\hat T_2, \hat T_3]=2i \hat T_2.
\eeq
The matrix representation of Eq. (\ref{compact}) is
\beq
\label{sisSU11}
\left(\begin{array}{c} 
\dot f_1 \\
\dot f_2   \\
\dot f_3 
\end{array} \right)=
\underbrace{\left(\begin{array}{ccc} 
-2f_3 &   0 & 2f_1  \\
0 &   2f_3&-2 f_2 \\
-f_2      &  f_1 & 0 
\end{array} \right)}_{=:\mathcal{G}}
\left(\begin{array}{c} 
h_1  \\
h_2\\
h_3
\end{array} \right).
\eeq
As in  the case of the SU(2) algebra the matrix $\mathcal{G}$ is not directly invertible. If the condition 
\beq
\label{consSU11}
\dot f_1f_2+\dot f_2f_1-2\dot f_3f_3=0
\eeq
or equivalently $f_1f_2-f_3^2=c$ holds, the previous system of equations becomes invertible and has infinite solutions
\beq
\label{Hs11}
h_a(t)=g_a(t)+\frac{f_a(t)}{f_3(t)}h_3(t)\quad a=1,2
\eeq
where $c$ is a constant, $g_1=-\dot f_1/(2f_3)$ and $g_2=\dot f_2/(2f_3)$, leaving $h_3(t)$ as an arbitrary free function of time.

For the example presented in Sec. \ref{sec:ho} of the expansion of a harmonic oscillator the generators $\hat T_a$ are represented by
 \beq
 \label{rep}
 \hat T_1=\frac{\hat p^2}{2}, \quad \hat T_2=\frac{\hat q^2}{2}, \quad \hat T_3=\frac{\hat p\hat q+\hat q\hat p}{2},
 \eeq
with $\hat q$ and $\hat p$ the position and momentum operators satisfying $[\hat q,\hat p]=i$. We partially fix the structure of $\hat H(t)$ 
\beq
\label{hamSU11}
\hat H(t)=\frac{\hat p^2}{2m}+\frac{m\omega^2(t)\hat q^2}{2}
\eeq
where $h_1(t)=1/m$ and $h_3(t)=0$ are imposed $\forall t$. The time dependency of $h_2(t)=m\omega^2(t)$ will be deduced to drive the system from a given Fock state $|n\rangle$ associated to $h_2(0)=m\omega_0^2$ to the
corresponding $|n\rangle$ state with $h_2(t_f)=m\omega_f^2$. In contrast with the TLS example, now a single control $h_2(t)$ will be designed. 
From the general formalism presented in \ref{app:general} a single invariant coefficient $f_a(t)$ is used. Using Eqs. (\ref{consSU11}) and (\ref{Hs11}) we can express 
$f_2=(\dot f_1^2+4h_1^2c)/(4h_1^2f_1)$ and $f_3=-\dot f_1/(2h_1)$ where $f_1$ satisfies
\beq
\label{f1eq}
\ddot f_1-\frac{\dot h_1}{h_1}\dot f_1-\frac{\dot f_1^2}{2f_1}-\frac{4ch_1^2}{2f_1}+2f_1h_1h_2=0,
\eeq
and the frictionless conditions $[\hat H(t_b),\hat I(t_b)]=0$ imposing $f_1(t_b)=\sqrt{ch_1(t_b)/h_2(t_b)}$ and $\dot f_1(t_b)=\ddot f_1(t_b)=0$.
This is just the Ermakov equation which is easily recognizable setting $h_1(t)=1/m, \dot h_1(t)=0, h_2(t)=m\omega(t)^2$, $c=m^2\omega_0^2$, and replacing $f_1=\rho^2$,
\beq
\label{rhoeq_app}
\ddot \rho+\omega^2(t)\rho=\frac{\omega_0^2}{\rho^3}.
\eeq
with $\rho(0)=1, \rho(t_f)=\sqrt{\omega_0/\omega_f}$ and $\dot\rho(t_b)=\ddot\rho(t_b)=0$ \cite{Chen2010}. Interpolating $f_1$ (or $\rho$) with at least six free parameters to be fixed by the frictionless conditions and solving Eq. (\ref{f1eq}) (or Eq. (\ref{rhoeq_app})) the
required control $\omega(t)$ is deduced.
In terms of $\rho$ the invariant associated with (\ref{hamSU11}) reads $\hat I(t)=\hat\pi^2/(2m)+m\omega_0^2\hat x^2/2$ where $[\hat x,\hat\pi]=i$ with $\hat x\equiv\hat q/\rho$ and $\hat\pi\equiv\rho\hat p-m\dot\rho\hat q$.
According to Eq. (\ref{wave}) the system at any time is,
\beqar
\label{Imode}
\langle q|\psi(t)\rangle&=&\sum_nc_ne^{-i\omega_0(n+1/2)\int_0^t\frac{dt'}{\rho^2}}\langle q|\phi_n(t)\rangle, \nonumber\\
\langle q|\phi_n(t)\rangle&=&\frac{e^{im\dot\rho q^2/(2\rho)}}{\sqrt{\rho}}\Phi_n(q/\rho),
\eeqar
where $\Phi_n(y)\equiv\langle y|n\rangle$ is the harmonic oscillator wave function composed by the Hermite polynomial with frequency $\omega_0$. Note that $\Phi_n(q/\rho)/\sqrt{\rho}$ represents the Fock state $|n\rangle$ in 
a harmonic trap of $\omega_0/\rho^2$ frequency.

The previous designed protocol is not only valid to connect single $|n\rangle$ to $|n\rangle$ Fock states but also coherent states \cite{Glauber1963}
\beq
|\alpha\rangle=e^{-|\alpha|^2/2}\sum_{n=0}^\infty\frac{\alpha^n}{\sqrt{n!}}|n\rangle
\eeq
forming a linear superposition. As at initial time $\hat H$ and $\hat I$ share a common basis $|\phi_n(0)\rangle=|n\rangle$ and according to Eqs. (\ref{wave}) and (\ref{Imode}) this initial state 
$|\psi(0)\rangle=|\alpha\rangle$ will evolve to \cite{Palmero2016}
\beq
\label{coh_state}
|\psi(t_f)\rangle=|\tilde\alpha\rangle=e^{-ig\omega_0/2}e^{-|\tilde\alpha|^2/2}\sum_{n=0}^{\infty}\frac{\tilde\alpha^n}{\sqrt{n!}}|\phi_n(t_f)\rangle,
\eeq
with $\tilde\alpha=\alpha e^{-ig\omega_0}$ and $g=\int_0^{t_f}dt'/\rho^2$. Thus, the system ends as a coherent state with frequency $\omega_f$.
\\
\bibliographystyle{iopart-num}

\begin{thebibliography}{10}
\expandafter\ifx\csname url\endcsname\relax
  \def\url#1{{\tt #1}}\fi
\expandafter\ifx\csname urlprefix\endcsname\relax\def\urlprefix{URL }\fi
\providecommand{\eprint}[2][]{\url{#2}}

\bibitem{koch16}
Koch C~P 2016 {\em Journal of Physics: Condensed Matter\/} {\bf 28} 213001

\bibitem{lidar03}
Lidar D~A and Birgitta~Whaley K 2003 {\em Decoherence-Free Subspaces and
  Subsystems\/} (Berlin, Heidelberg: Springer Berlin Heidelberg) pp 83--120

\bibitem{wisemanbook}
Wiseman H and Milburn G 2010 {\em Quantum Measurement and Control\/} (Cambridge
  University Press) ISBN 9780521804424

\bibitem{lloyd99}
Viola L, Knill E and Lloyd S 1999 {\em Phys. Rev. Lett.\/} {\bf 82}(12)
  2417--2421

\bibitem{Chen2010}
Chen X, Ruschhaupt A, Schmidt S, del Campo A, Gu\'ery-Odelin D and Muga J~G
  2010 {\em Phys. Rev. Lett.\/} {\bf 104}(6) 063002

\bibitem{torrontegui13}
Torrontegui E, Ib{\'a}{\~n}ez S, Mart{\'\i}nez-Garaot S, Modugno M, del Campo
  A, Gu{\'e}ry-Odelin D, Ruschhaupt A, Chen X and Muga J~G 2013 {\em Adv. At.
  Mol. Opt. Phys\/} {\bf 62} 117--169

\bibitem{levy12}
Levy A and Kosloff R 2012 {\em Phys. Rev. Lett.\/} {\bf 108}(7) 070604

\bibitem{levy17}
Levy A, Torrontegui E and Kosloff R 2017 {\em Phys. Rev. A\/} {\bf 96}(3)
  033417

\bibitem{koch15b}
Reich D~M, Katz N and Koch C~P 2015 {\em Scientific reports\/} {\bf 5}

\bibitem{khasin11}
Khasin M and Kosloff R 2011 {\em Physical review letters\/} {\bf 106} 123002

\bibitem{sarandy11}
Sarandy M, Duzzioni E and Serra R 2011 {\em Physics Letters A\/} {\bf 375}
  3343--3347

\bibitem{Palmero2017}
Palmero M, Mart{\'{i}}nez-Garaot S, Leibfried D, Wineland D~J and Muga J~G 2017
  {\em Physical Review A\/} {\bf 95} 022328 ISSN 2469-9926

\bibitem{Onofrio2017}
Onofrio R 2017 {\em Physics-Uspekhi\/} {\bf 59} 1129

\bibitem{Torrontegui2011}
Torrontegui E, Ib\'a\~nez S, Chen X, Ruschhaupt A, Gu\'ery-Odelin D and Muga
  J~G 2011 {\em Phys. Rev. A\/} {\bf 83}(1) 013415

\bibitem{bowler}
Bowler R, Gaebler J, Lin Y, Tan T~R, Hanneke D, Jost J~D, Home J~P, Leibfried D
  and Wineland D~J 2012 {\em Phys. Rev. Lett.\/} {\bf 109}(8) 080502

\bibitem{muga12b}
Chen X, Lizuain I, Ruschhaupt A, Gu\'ery-Odelin D and Muga J~G 2010 {\em Phys.
  Rev. Lett.\/} {\bf 105}(12) 123003

\bibitem{2levelEXP1}
Bason M~G, Viteau M, Malossi N, Huillery P, Arimondo E, Ciampini D, Fazio R,
  Giovannetti V, Mannella R and Morsch O 2012 {\em Nature Physics\/} {\bf 8}
  147--152

\bibitem{2levelEXP2}
Zhang J, Shim J~H, Niemeyer I, Taniguchi T, Teraji T, Abe H, Onoda S, Yamamoto
  T, Ohshima T, Isoya J and Suter D 2013 {\em Phys. Rev. Lett.\/} {\bf 110}(24)
  240501

\bibitem{Zhou2017_330}
Zhou B~B, Baksic A, Ribeiro H, Yale C~G, Heremans F~J, Jerger P~C, Auer A,
  Burkard G, Clerk A~A and Awschalom D~D 2017 {\em Nature Physics\/} {\bf 13}
  330--334 ISSN 1745-2473

\bibitem{Torrontegui2012}
Torrontegui E, Chen X, Modugno M, Schmidt S, Ruschhaupt A and Muga J~G {\em New
  Journal of Physics\/} {\bf 14} 013031

\bibitem{sta_exp1}
Schaff J~F, Song X~L, Vignolo P and Labeyrie G 2010 {\em Phys. Rev. A\/} {\bf
  82}(3) 033430

\bibitem{schaff11}
Schaff J~F, Capuzzi P, Labeyrie G and Vignolo P 2011 {\em New Journal of
  Physics\/} {\bf 13} 113017

\bibitem{split_sta}
Torrontegui E, Mart\'{\i}nez-Garaot S, Modugno M, Chen X and Muga J~G 2013 {\em
  Phys. Rev. A\/} {\bf 87}(3) 033630

\bibitem{Kiely2016}
Kiely A, Benseny A, Busch T and Ruschhaupt A 2016 {\em Journal of Physics B:
  Atomic, Molecular and Optical Physics\/} {\bf 49} 215003 ISSN 0953-4075

\bibitem{adolfo_multi}
del Campo A 2013 {\em Phys. Rev. Lett.\/} {\bf 111}(10) 100502

\bibitem{SebCriAdo}
Deffner S, Jarzynski C and del Campo A 2014 {\em Phys. Rev. X\/} {\bf 4}(2)
  021013

\bibitem{Takahashi2013}
Takahashi K 2013 {\em Phys. Rev. E\/} {\bf 87}(6) 062117

\bibitem{Martinez-Garaot2015_043406}
Mart{\'{i}}nez-Garaot S, Ruschhaupt A, Gillet J, Busch T and Muga J~G 2015 {\em
  Physical Review A\/} {\bf 92} 043406 ISSN 1050-2947

\bibitem{Masuda2015}
Masuda S and Rice S~A 2015 {\em The Journal of Physical Chemistry A\/} {\bf
  119} 3479--3487 ISSN 1089-5639

\bibitem{Tseng2012}
Tseng S~Y and Chen X 2012 {\em Optics Letters\/} {\bf 37} 5118 ISSN 0146-9592

\bibitem{Martinez-Garaot2014_2306}
Mart{\'{i}}nez-Garaot S, Tseng S~Y and Muga J~G 2014 {\em Optics Letters\/}
  {\bf 39} 2306 ISSN 0146-9592

\bibitem{Longhi2017}
Longhi S 2017 {\em Phys. Rev. A\/} {\bf 95}(6) 062122

\bibitem{Torrontegui2017}
Torrontegui E, Lizuain I, Gonz{\'{a}}lez-Resines S, Tobalina A, Ruschhaupt A,
  Kosloff R and Muga J~G 2017 {\em Physical Review A\/} {\bf 96} 022133 ISSN
  2469-9926

\bibitem{Gonzalez2017}
Gonz\'alez-Resines S, Gu\'ery-Odelin D, Tobalina A, Lizuain I, Torrontegui E
  and Muga J~G 2017 {\em Phys. Rev. Applied\/} {\bf 8}(5) 054008

\bibitem{Chen2014}
Chen Y~H, Xia Y, Chen Q~Q and Song J 2014 {\em Laser Physics Letters\/} {\bf
  11} 115201

\bibitem{Chen2015_012325}
Chen Y~H, Xia Y, Chen Q~Q and Song J 2015 {\em Phys. Rev. A\/} {\bf 91}(1)
  012325

\bibitem{Ruschhaupt2012}
Ruschhaupt A, Chen X, Alonso D and Muga J~G 2012 {\em New Journal of Physics\/}
  {\bf 14} 093040 ISSN 1367-2630

\bibitem{Lu2013}
Lu X~J, Chen X, Ruschhaupt A, Alonso D, Gu\'erin S and Muga J~G 2013 {\em Phys.
  Rev. A\/} {\bf 88}(3) 033406

\bibitem{Daems2013}
Daems D, Ruschhaupt A, Sugny D and Gu{\'{e}}rin S 2013 {\em Physical Review
  Letters\/} {\bf 111} 050404 ISSN 0031-9007

\bibitem{Kiely2014}
Kiely A and Ruschhaupt A 2014 {\em Journal of Physics B: Atomic, Molecular and
  Optical Physics\/} {\bf 47} 115501 ISSN 0953-4075

\bibitem{Lu2014_063414}
Lu X~J, Muga J~G, Chen X, Poschinger U~G, Schmidt-Kaler F and Ruschhaupt A 2014
  {\em Physical Review A\/} {\bf 89} 063414 ISSN 1050-2947

\bibitem{Wu2015}
Wu S~L, Zhang X~Y and Yi X~X 2015 {\em Phys. Rev. A\/} {\bf 92}(6) 062122

\bibitem{Wu2017_042104}
Wu S~L, Huang X~L, Li H and Yi X~X 2017 {\em Phys. Rev. A\/} {\bf 96}(4) 042104

\bibitem{Sarandy2007}
Sarandy M~S, Duzzioni E~I and Moussa M~H~Y 2007 {\em Phys. Rev. A\/} {\bf
  76}(5) 052112

\bibitem{Maamache2017}
Maamache M, Kaltoum~Djeghiour O, Mana N and Koussa W 2017 {\em The European
  Physical Journal Plus\/} {\bf 132} 383 ISSN 2190-5444

\bibitem{Luo2015}
Luo D~W, Pyshkin P~V, Lam C~H, Yu T, Lin H~Q, You J~Q and Wu L~A 2015 {\em
  Phys. Rev. A\/} {\bf 92}(6) 062127

\bibitem{Kiely2017}
Kiely A, Muga J~G and Ruschhaupt A 2017 {\em Phys. Rev. A\/} {\bf 95}(1) 012115

\bibitem{rabitz88}
Peirce A~P, Dahleh M~A and Rabitz H 1988 {\em Phys. Rev. A\/} {\bf 37}(12)
  4950--4964

\bibitem{koch15}
Glaser S~J, Boscain U, Calarco T, Koch C~P, Köckenberger W, Kosloff R, Kuprov
  I, Luy B, Schirmer S, Schulte-Herbrüggen T, Sugny D and Wilhelm F~K 2015
  {\em The European Physical Journal D\/} {\bf 69} 279 ISSN 1434-6060

\bibitem{gorini76}
Gorini V and Kossakowski A 1976 {\em J. Math. Phys.\/} {\bf 17} 1298

\bibitem{milburn91}
Milburn G~J 1991 {\em Phys. Rev. A\/} {\bf 44}(9) 5401--5406

\bibitem{kallush14}
Kallush S, Khasin M and Kosloff R 2014 {\em New Journal of Physics\/} {\bf 16}
  015008

\bibitem{diosi88}
Di\'{o}si L 1988 {\em Physics Letters A\/} {\bf 129} 419--423

\bibitem{Luczka1991}
Luczka J and Niemiec M 1991 {\em Journal of Physics A: Mathematical and
  General\/} {\bf 24} L1021

\bibitem{levy14a}
Kosloff R and Levy A 2014 {\em Annual Review of Physical Chemistry\/} {\bf 65}
  365

\bibitem{breuer}
{H-P Breuer and F Petruccione} 2002 {\em Open quantum systems\/} (Oxford
  university press)

\bibitem{luczka91}
{J Luczka} 1991 {\em Czechoslovak Journal of Physics\/} {\bf 41} {289}

\bibitem{Lewis1969}
Jr H~R~L and Riesenfeld W~B 1969 {\em Journal of Mathematical Physics\/} {\bf
  10} 1458--1473

\bibitem{muga12}
Ruschhaupt A, Chen X, Alonso D and Muga J~G 2012 {\em New Journal of Physics\/}
  {\bf 14} 093040

\bibitem{demirplak03}
Demirplak M and Rice S~A 2003 {\em The Journal of Physical Chemistry A\/} {\bf
  107} 9937--9945

\bibitem{berry09}
Berry M~V 2009 {\em Journal of Physics A: Mathematical and Theoretical\/} {\bf
  42} 365303

\bibitem{rabitz08}
Pechen A, Prokhorenko D, Wu R and Rabitz H 2008 {\em Journal of Physics A:
  Mathematical and Theoretical\/} {\bf 41} 045205

\bibitem{durt08}
DURT T, ENGLERT B~G, BENGTSSON I and ŻYCZKOWSKI K 2010 {\em International
  Journal of Quantum Information\/} {\bf 08} 535--640

\bibitem{Diaz2012}
Juli\'a-D\'{\i}az B, Torrontegui E, Martorell J, Muga J~G and Polls A 2012 {\em
  Phys. Rev. A\/} {\bf 86}(6) 063623

\bibitem{Yuste2013}
Yuste A, Juli\'a-D\'{\i}az B, Torrontegui E, Martorell J, Muga J~G and Polls A
  2013 {\em Phys. Rev. A\/} {\bf 88}(4) 043647

\bibitem{banchi15}
Banchi L, Braunstein S~L and Pirandola S 2015 {\em Phys. Rev. Lett.\/} {\bf
  115}(26) 260501

\bibitem{Torrontegui2014}
Torrontegui E, Mart\'{\i}nez-Garaot S and Muga J~G 2014 {\em Phys. Rev. A\/}
  {\bf 89}(4) 043408

\bibitem{Kaushal1981}
Kaushal R~S and Korsch H~J 1981 {\em Journal of Mathematical Physics\/} {\bf
  22} 1904--1908

\bibitem{lewis69}
Lewis~Jr H~R and Riesenfeld W 1969 {\em Journal of Mathematical Physics\/} {\bf
  10} 1458--1473

\bibitem{Maamache1995}
Maamache M 1995 {\em Phys. Rev. A\/} {\bf 52}(2) 936--940

\bibitem{Kaushal1993}
Kaushal R~S and Mishra S~C 1993 {\em Journal of Mathematical Physics\/} {\bf
  34} 5843--5850

\bibitem{Monteoliva1994}
Monteoliva D~B, Korsch H~J and Nunez J~A 1994 {\em Journal of Physics A:
  Mathematical and General\/} {\bf 27} 6897

\bibitem{Chen2011}
Chen X, Torrontegui E and Muga J~G 2011 {\em Phys. Rev. A\/} {\bf 83}(6) 062116

\bibitem{Glauber1963}
Glauber R~J 1963 {\em Phys. Rev.\/} {\bf 131}(6) 2766--2788

\bibitem{Palmero2016}
Palmero M, Wang S, Gu\'ery-Odelin D, Li J~S and Muga J~G 2016 {\em New Journal
  of Physics\/} {\bf 18} 043014

\end{thebibliography}
\providecommand{\newblock}{}

\end{document}